\documentclass[twocolumn,tighten]{aastex62}

\usepackage{xcolor}
\usepackage{hyperref}
\usepackage[caption=false]{subfig}
\usepackage{longtable}
\usepackage{afterpage}
\usepackage{xspace}
\usepackage{upgreek}
\usepackage{multirow}

\graphicspath{{./}{figures/}}

\received{June 18, 2019}
\revised{November 20, 2019}
\accepted{December 22, 2019}

\shorttitle{Breakthrough Listen 1.10--3.45 GHz observations}
\shortauthors{D. C. Price et. al.}

\begin{document}

\newcommand{\tseti}{\textsc{\emph{turbo}SETI} }
\newcommand{\tiddalik}{\textsc{Tiddalik} }
\newcommand{\blimpy}{\textsc{Blimpy} }
\newcommand{\Hzs}{Hz\,s$^{-1}$}
\newcommand{\EventBW}{$\Delta\nu_{\rm{event}}$}

\newcommand{\nstar}{{\color{black}1327 }}
\newcommand{\nstaradd}{{\color{black}641 }}  

\newcommand{\nobsgbtS}{{\color{black}6456 }}
\newcommand{\nobsgbtL}{{\color{black}6042 }}
\newcommand{\nobsgbtSL}{{\color{black}12504 }}

\newcommand{\nstargbt}{{\color{black}1138 }}
\newcommand{\nstargbtS}{{\color{black}1005 }}
\newcommand{\nstargbtL}{{\color{black}882 }}
\newcommand{\nstargbtSL}{{\color{black}749 }}

\newcommand{\ncadencegbt}{{\color{black}2089 }}
\newcommand{\ncadencegbtS}{{\color{black}1076 }}
\newcommand{\ncadencegbtL}{{\color{black}1013 }}

\newcommand{\nhrpks}{483.0}
\newcommand{\nhrgbtL}{506.5}
\newcommand{\nhrgbtS}{538.0}

\newcommand{\nobspks}{{\color{black}5796}\xspace}
\newcommand{\nstarpks}{{\color{black}189}\xspace}
\newcommand{\nstarpksIsaacson}{{\color{black}183}\xspace}
\newcommand{\nstarpksExtra}{{\color{black}6}\xspace}
\newcommand{\ncadencepksdb}{{\color{black}990}\xspace}
\newcommand{\ncadencepks}{{\color{black}966}\xspace}

\newcommand{\nevent}{{\color{black} zero candidates}\xspace}
\newcommand{\neventpks}{{\color{black}77\xspace}}
\newcommand{\neventgbtL}{{\color{black}15998}\xspace}
\newcommand{\neventgbtS}{{\color{black}5102}\xspace}

\newcommand{\neventpksnstar}{20\xspace}  
\newcommand{\neventgbtLnstar}{831\xspace}
\newcommand{\neventgbtSnstar}{511\xspace}

\newcommand{\ngroupspks}{{\color{black}60}\xspace}
\newcommand{\ngroupsgbtL}{{\color{black}4522}\xspace}
\newcommand{\ngroupsgbtS}{{\color{black}1572}\xspace}

\newcommand{\nfinalpks}{0\xspace}
\newcommand{\nfinalL}{0\xspace}
\newcommand{\nfinalS}{0\xspace}

\newcommand{\nhitspks}{4.45M\xspace}
\newcommand{\nhitsgbtL}{{\color{black}37.14M}\xspace}
\newcommand{\nhitsgbtS}{{\color{black}10.12M}\xspace}

\newcommand{\nhitsgbtLzerodrift}{21.90M\xspace}
\newcommand{\nhitsgbtLnegativedrift}{13.9M\xspace}
\newcommand{\nhitsgbtLpositivedrift}{1.37M\xspace}

\newcommand{\nhitsgbtSzerodrift}{7.36M\xspace}
\newcommand{\nhitsgbtSpositivedrift}{1.21M\xspace}
\newcommand{\nhitsgbtSnegativedrift}{1.55M\xspace}


\newcommand{\BI}{\textit{Breakthrough Initiatives} }
\newcommand{\BLI}{\textit{Breakthrough Listen Initiative} }
\newcommand{\BL}{BL\xspace}

\title{The Breakthrough Listen Search for Intelligent Life: \\ 
 Observations of \nstar Nearby Stars over 1.10--3.45\,GHz}


\newcommand{\UCB}{Department of Astronomy,  University of California Berkeley, Berkeley CA 94720}
\newcommand{\SSL}{Space Sciences Laboratory, University of California, Berkeley, Berkeley CA 94720}
\newcommand{\SWIN}{Centre for Astrophysics \& Supercomputing, Swinburne University of Technology, Hawthorn, VIC 3122, Australia}
\newcommand{\GBT}{Green Bank Observatory,  West Virginia, 24944, USA}
\newcommand{\OXF}{Astronomy Department, University of Oxford, Keble Rd, Oxford, OX13RH, United Kingdom}
\newcommand{\NIJ}{Department of Astrophysics/IMAPP,Radboud University, Nijmegen, Netherlands}
\newcommand{\ATNF}{Australia Telescope National Facility, CSIRO, PO Box 76, Epping, NSW 1710, Australia}
\newcommand{\HOU}{Hellenic Open University, School of Science \& Technology, Parodos Aristotelous, Perivola Patron, Greece}

 \newcommand{\USQ}{University of Southern Queensland, Toowoomba, QLD 4350, Australia}

\newcommand{\SETI}{SETI Institute, Mountain View, California}
\newcommand{\KZA}{University of Malta, Institute of Space Sciences and Astronomy}
\newcommand{\PWJD}{The Breakthrough Initiatives, NASA Research Park, Bld. 18, Moffett Field, CA, 94035, USA}

\newcommand{\mgo}{M\&P }
\newcommand{\refsec}[1]{Sec.~\ref{#1}}
\newcommand{\reffig}[1]{Fig.~\ref{#1}}
\newcommand{\reftab}[1]{Tab.~\ref{#1}}

\newcommand{\ee}[1]{\textbf{\color{blue} EE: #1}} 
\newcommand{\GSF}[1]{\textbf{\color{red} GSF: #1}} 
\newcommand{\dc}[1]{\textbf{\color{orange} DC: #1}} 
\newcommand{\dcp}[1]{\textbf{\color{purple} DCP: #1}} 
\newcommand{\todo}[1]{\textbf{\color{purple} TODO: #1}} 
\newcommand{\bcl}[1]{\textbf{\color[rgb]{0.2,0,0.5} BCL: #1}} 
\newcommand{\vg}[1]{\textbf{\color{green} VG: #1}} 
\newcommand{\ngi}[1]{\textbf{\color{cyan} NG: #1}} 
\newcommand{\diff}[1]{{\color{black} #1}} 

\newcommand{\apvs}[1]{\textbf{\color{green} APVS: #1}} 

\correspondingauthor{Danny C. Price}
\email{dancpr@berkeley.edu}

\author[0000-0003-2783-1608]{Danny C.\ Price}
\affiliation{\UCB}
\affiliation{\SWIN}

\author[0000-0003-2516-3546]{J. Emilio Enriquez}
\affiliation{\UCB}
\affiliation{\NIJ}

\author[0000-0002-7461-107X]{Bryan Brzycki}
\affiliation{\UCB}

\author[0000-0003-4823-129X]{Steve Croft}
\affiliation{\UCB}

\author[0000-0002-8071-6011]{Daniel Czech}
\affiliation{\UCB}

\author[0000-0003-3197-2294]{David DeBoer}
\affiliation{\UCB}

\author{Julia DeMarines}
\affiliation{\UCB}

\author[0000-0002-7559-4291]{Griffin Foster}
\affiliation{\UCB}
\affiliation{\OXF}

\author[0000-0002-8604-106X]{Vishal Gajjar}
\affiliation{\UCB}

\author[0000-0002-9787-1149]{Nectaria Gizani}
\affiliation{\UCB}
\affiliation{\HOU}

\author[0000-0002-8191-3885]{Greg Hellbourg}
\affiliation{\UCB}

\author[0000-0002-0531-1073]{Howard Isaacson}
\affiliation{\UCB}
\affiliation{\USQ}

\author{Brian Lacki}
\affiliation{Breakthrough Listen, \UCB}

\author{Matt Lebofsky}
\affiliation{\UCB}

\author{David H.\ E.\ MacMahon}
\affiliation{\UCB}

\author{Imke de Pater}
\affiliation{\UCB}

\author[0000-0003-2828-7720]{Andrew P.\ V.\ Siemion}
\affiliation{\UCB}
\affiliation{\SETI}
\affiliation{\NIJ}
\affiliation{\KZA}

\author{Dan Werthimer}
\affiliation{\UCB}

\author[0000-0002-2670-188X]{James A.\ Green}
\affiliation{\ATNF}

\author[0000-0003-4810-7803]{Jane F.\ Kaczmarek}
\affiliation{\ATNF}

\author[0000-0002-2660-9581]{Ronald J.\ Maddalena}
\affiliation{\GBT}

\author[0000-0002-9332-8616]{Stacy Mader}
\affiliation{\ATNF}

\author{Jamie Drew}
\affiliation{\PWJD}

\author{S. Pete Worden}
\affiliation{\PWJD}

\begin{abstract}
Breakthrough Listen (BL) is a ten-year initiative to search for signatures of technologically capable life beyond Earth via radio and optical observations of the local Universe. A core part of the BL program is a comprehensive survey of 1702 nearby stars at radio wavelengths (1--10~GHz). Here, we report on observations with the 64-m CSIRO Parkes radio telescope in New South Wales, Australia, and the 100-m Robert C. Byrd Green Bank radio telescope in West Virginia, USA. 
Over 2016 January to 2019 March, a sample of \nstargbt \,stars was observed at Green Bank using the 1.10--1.90\,GHz and 1.80--2.80~GHz receivers, and \nstarpks stars were observed with Parkes over 2.60--3.45~GHz. We searched these data for the presence of engineered signals with Doppler-acceleration drift rates between $\pm$4\,\Hzs. Here, we detail our data analysis techniques and provide examples of detected events. After excluding events with characteristics consistent with terrestrial radio interference, we are left with zero candidates. \diff{That is, we find no evidence of putative radio transmitters above $2.1\times10^{12}$\,W, and $9.1\times10^{12}$\,W for Green Bank and Parkes observations, respectively.} These observations constitute the most comprehensive search over 1.10--3.45\,GHz for technosignatures to date. All data products, totalling $\sim$219\,TB, are available for download as part of the first \BL data release (DR1), as described in a companion paper (Lebofsky et. al., 2019)
\end{abstract}

\keywords{astrobiology --- technosignature --- SETI --- extraterrestrial intelligence}


\section{Introduction} \label{sec:intro}

If we are to detect life beyond Earth in the next few decades, it will be by one of three ongoing efforts. We may find life in the Solar system by physically examining the environment of our planetary neighbors and their moons. Optical spectroscopy may detect biosignatures in the atmospheres of nearby exoplanets, indicating the presence of life. Or, we may detect evidence of advanced life via technosignatures: signals of engineering that are discernable from astrophysical processes.

These three methods are complementary, as they probe different manifestations of life at different distance scales and time scales in life evolution. 
The latter approach is known as the Search for Extraterrestrial Intelligence (SETI), and is the only method that can conceivably detect signals from \diff{beyond the nearest stars} with current or near-term technology. SETI seeks not just to detect signs of life, but also to constrain the probability of the emergence of intelligence life: whether we are the sole inhabitants of the Universe, or whether it is ours to share. 

Radio searches for extraterrestrial intelligence (ETI) have been ongoing since the 1960s \citep{Drake1961}. The sensitivity and speed of SETI searches is intimately tied to our own technological capability; as technology progresses, so too do the capabilities and sensitivity of radio telescopes. Of particular importance in this regard are the digital instruments used in radio SETI searches. The instantaneous bandwidth of these systems have expanded from hundreds of hertz \citep{Drake1961}, to kilohertz \citep[e.g.][]{Werthimer1985,Horowitz:1993}, to megahertz \citep{Werthimer1995, Tarter1996}, and through to tens of gigahertz \citep{Macmahon:2017}---a factor of $10^8$---over the course of roughly 40 years. The search has also expanded from single dish radio telescopes to interferometers \citep[e.g.][]{Welch2009, Rampadarath:2012, Tingay:2016, Gray:2017dr, Tingay:2018a, Tingay:2018b}, optical \citep{Howard2004, Tellis:2015, WrightS2001, Reines2002, Stone2005, Howard2007} and infrared wavelengths \citep{Slysh1985, Carrigan2009, WrightJ2014a, WrightS:2015, Griffith2015}. In tandem with increased frequency coverage, the sensitivity and field of view of telescopes continue to increase, allowing ever more exquisite measurements to be made. 

\subsection{Breakthrough Listen}

The Breakthrough Listen (BL) initiative represents the current state-of-the-art for SETI search strategies and capabilities. \BL is a ten-year initiative to search for technosignatures at radio and optical wavelengths \citep{Worden:2017}. Having commenced observations in 2016, the program expands the capability of existing telescopes for SETI observations by installing wide-bandwidth data recording and analysis systems capable of recording raw digitizer samples direct to disk \citep[see][]{Macmahon:2017, Price:2018, Lebofsky:2019}. In its initial years, BL is conducting observations with the 100-m Robert C. Byrd Green Bank (henceforth GBT) radio telescope in West Virginia, USA; and the 64-m CSIRO Parkes radio telescope in New South Wales, Australia. New digital systems have been installed at both telescopes to allow for capture of voltage data across the full bandwidth provided by the receivers of the two telescopes \citep{Macmahon:2017, Price:2018}. At optical wavelengths, the Automated Planet Finder telescope in California, USA, \citep{Radovan2014} is conducting a search for laser lines in high-resolution spectra \citep[e.g.][]{Lipman:2019}. 
\BL is currently conducting a survey of several thousand nearby stars, 100 nearby galaxies, and the Galactic plane; further details may be found in \citet{Isaacson:2017}. An initial analysis of 692 stars, over 1.1--1.9~GHz, is presented in \citet{Enriquez:2017}; no high-duty-cycle narrowband radio transmissions with equivalent isotropic radiated power (EIRP) of $>$10$^{13}$\,W  were found in this sample. 

Here, we present an analysis of \nstar star targets taken over 1.10--3.45\,GHz (L-band and S-band), including reanalysis of the observations of 692 stars detailed in \citet{Enriquez:2017}. In addition to covering greater bandwidth, we improve on the \citet{Enriquez:2017} limits by using a lower signal to noise (S/N) cutoff (10 vs. 25), a larger range of drift rates ($\pm$4\,Hz s$^{-1}$ vs. $\pm$2\,Hz s$^{-1}$) and enhanced signal identification logic. 

This paper is organized as follows. In \refsec{sec:obs}, a summary of observations is given. Data analysis strategies are detailed in \refsec{sec:analysis}. Results are given in \refsec{sec:results}, followed by discussion and conclusions.
An explanation of the data formats and archiving strategy is given in a companion paper \citep{Lebofsky:2019}.

\section{Observations} \label{sec:obs}

We used the Green Bank and Parkes telescopes to observe nearby stars at frequencies between 1.10--3.45\,GHz (see \reftab{tab:receivers}). This section provides a summary of the stellar targets observed, details of the two telescopes used, observational details, and an overview of data products. A full listing of observed stars can be found at \href{https://seti.berkeley.edu/listen2019}{seti.berkeley.edu/listen2019}.

\begin{table}
\begin{center}
\caption{\label{tab:receivers} Details of the receivers used here. 
}
\begin{tabular}{llccc}
\hline

Telescope & Receiver & Frequency & $T_{\rm{sys}}$ & SEFD \\
          &          &   (GHz)   &    (K)         & (Jy) \\
\hline
\hline
Green Bank & L-band & 1.10--1.90  &  25   & 10 \\
Green Bank & S-band & 1.80--2.80  &  25   & 10 \\
Parkes     & 10-cm  & 2.60--3.45  &  35   & 34 \\
\hline
\end{tabular}
\end{center}
\end{table}

\subsection{Star sample}

\begin{table*}\label{tab:extra-stars}
\begin{center}
\caption{\label{tab:pks-extra} Stars within 5 pc added to the \citet{Isaacson:2017} sample for improved volume completeness below -15$^\circ$ declination.}

\begin{tabular}{lcllccr}
\hline
Star &   Epoch & RA &  DEC &   Distance (pc) & Spectral Type & Reference \\
\hline
\hline
    DENIS J025503.3-470049 & 2000 &        2:55:03.6 &      -47:00:51.0 & 4.9 & L8/L9 & \cite{2003yCat.2246....0C} \\ 


           SCR J1845-6357A & 2000 &       18:45:05.3 &      -63:57:47.1 & 3.9 & M8.5 & \cite{2006ApJ...641L.141B} \\

  WISE J035000.32-565830.2 & 2000 &        3:50:00.3 &      -56:58:30.2 & 5.4 & Y1 & \cite{2012yCat.2311....0C} \\
  
  WISE J053516.80-750024.9 & 2000 &        5:35:16.8 &      -75:00:24.9 & 2.0 & Y1 & \cite{2012yCat.2311....0C} \\
  
  WISE J104915.57-531906.1 & 2000 &       10:49:18.9 &      -53:19:10.1 & 2.0 & L7.5, T0.5 & \cite{2003yCat.2246....0C} \\
  
 WISEA J154045.67-510139.3 & 2000 &       15:40:45.7 &      -51:01:39.3 & 4.6 & M7 & \cite{2014ApJ...783..122K} \\

\hline
\end{tabular}
\end{center}
\end{table*}

We observed nearby stellar targets chosen from the \citet{Isaacson:2017} 1702-star sample, with the Green Bank and Parkes radio telescopes. The \citet{Isaacson:2017} sample is comprised of stars selected from the RECONS and Gliese catalogs of nearby stars \citep{Gliese1995}, and the well-characterized \emph{Hipparcos} catalog \citep{Perryman1997}. The sample is constructed to contain all stars within 5\,pc in the Gliese and RECONS catalogs, and a broad sampling of main sequence stars within 5--50\,pc from \emph{Hipparcos}. 

 For observations with Parkes, the \cite{Isaacson:2017} sample was augmented with a small number of recently discovered brown dwarfs and other stars within 5\,pc, below a declination of -15$^\circ$; these are detailed in \reftab{tab:pks-extra}.  In total, \nstar distinct primary `A' star targets were observed. It should be noted that this is still not a complete list, and the continuing discovery of nearby low-mass stars necessitates periodic revisiting of our volume-limited sample. Also we note that not all stars were observed with each single receiver. The number of stars per receiver are broken down below and summarized in \reftab{tab:observations}. 

\subsection{Observing Strategy}

At both GBT and Parkes, we employ an observing strategy whereby a target is observed for five minutes (`ON' source), then a reference location is observed for five minutes (`OFF' source). This ON-OFF strategy is repeated three times for each target, taking a total of 30 minutes (plus slew time). This strategy is used to allow for discrimination of bonafide signals of interest from radio interference (RFI): any signal that appears in both ON and OFF pointings at power levels inconsistent with the known off-axis gain of the telescope is considered RFI. To further discriminate RFI-induced false positives, we enforce that signals must appear in all three ON pointings, or in other words that the signals are continuous throughout the observation. Further discrimination may then be done by enforcing that signals exhibit a non-zero doppler acceleration (see \refsec{sec:analysis}), and by cross-reference between observations. 

Following \citet{Enriquez:2017}, we refer to a strategy of  applying a constant offset for OFF positions as ABABAB; the strategy of interspersing  three different nearby `secondary' targets is referred to as ABACAD. At Parkes, an ABABAB strategy is used with a fixed 0.5$^\circ$ declination offset ($\sim$ three FWHM beamwidths at the lowest observing frequency); at GBT an ABACAD strategy is predominantly used. For GBT ABACAD observations, nearby stars selected from the \emph{Hipparcos} catalog are used for OFF-source pointings \diff{(these `secondary' stars may be searched for technosignature or flare activity at a later date)}.
The \emph{Hipparcos} stars are chosen to be between 1.2 and 3.6 degrees away from the primary target ($>8$ FWHM beamwidths at the lowest observing frequency). This separation was chosen to be sufficiently far from the primary beam, within a reasonable slew time and encompassing an area that likely holds 3 \emph{Hipparcos} stars. A search of the secondary targets is also possible, but outside the scope of this paper.

\subsection{Green Bank Telescope}

\begin{table}
\begin{center}
\caption{\label{tab:observations} Summary of GBT S, GBT L, and Parkes 10-cm observations.
}
\begin{tabular}{lccc}
\hline

Receiver & No. cadences & No. targets  & Hours \\
\hline
\hline
GBT L        & \ncadencegbtL  & \nstargbtL  & \nhrgbtL  \\
GBT S        & \ncadencegbtS  & \nstargbtS  & \nhrgbtS \\
Parkes 10-cm & \ncadencepks   & \nstarpks   & \nhrpks  \\
\hline
\end{tabular}
\end{center}
\end{table}

The GBT is a 100-m radio telescope located in West Virginia, USA (38$^\circ$25'59.236''s N,79$^\circ$50'23.406'' W), operated by the Green Bank Observatory. The telescope is located within a federally protected `Radio Quiet' zone, in which most radio transmissions are prohibited to minimize radio frequency interference (RFI). Approximately 20\% of the annual observing time for the GBT is dedicated to \BL. The GBT has an operational range of 0.3--110\,GHz, depending on the receiver equipped during observation. For the analyses detailed here, we used the 1.10--1.90\,GHz (L-band) receiver, and 1.80--2.80\,GHz (S-band)   receiver, both with a system temperature of 20\,K, resulting in a system equivalent flux density (SEFD) of 10\,Jy. The L-band contains a user-selectable notch filter band between 1.20 and 1.34\,GHz (which is always used in \BL observations) and the S-band contains a permanently installed superconducting notch filter band between 2.3 and 2.36\,GHz \citep{gbtpropguide}.

At the time of writing, the nearby star observation program at GBT is currently focusing on observations with the 4.0--7.8\,GHz (C-band) and 7.8--12.3\,GHz (X-band) receivers. Completion of these programs are at about 80\% and 60\% respectively. Use of the 18--27.5\,GHz (KFPA-band) receiver for \BL observations is being commissioned, with the full 10\,GHz bandwidth of raw voltage data produced by this receiver available to the current BL backend \citep{Macmahon:2017}. Analysis of these data is expected to be included in future publications.

A total of \nobsgbtSL 5-minute observations with the GBT ($\sim$1044 hr) are used for this work, conducted over the period 2016 January 1 to 2019 March 23 (MJD 57388--58565), summarised in \reftab{tab:observations}. Out of these observations, \nobsgbtL were carried out with the L-band receiver (\ncadencegbtL cadences), and \nobsgbtS were carried out with the S-band receiver (\ncadencegbtS cadences). Due to a small number of repeated observations, where the star was selected more than once from the target database, a total \nstargbt of the primary `A' stars were observed: \nstargbtSL at both bands, \nstargbtL at L-band, and \nstargbtS at S-band.  

\subsection{Parkes Radio Telescope}

The CSIRO Parkes radio telescope is a 64-m telescope located in New South Wales, Australia (32$^\circ$59'59.8''S, 148$^\circ$15'44.3''E). As with the GBT, Parkes is equipped with a suite of receivers, which cover 0.6--26.0\,GHz. Over the period October 2016 to September 2021, a quarter of the annual observation time of the Parkes 64-m radio telescope has been dedicated to the BL program. The analyses detailed here are from data taken with 10-cm component of the Parkes `10-50' receiver, which covers 2.60--3.45\,GHz. This receiver has a nominal system temperature of 35\,K, with a corresponding SEFD of 34\,Jy. 

In contrast to the GBT, the Parkes observations of the nearby star sample used herein include multiple epochs. This was motivated primarily by technical concerns: firstly, the Parkes system was deployed in stages over a period of several months when the receiver availability varied (two receivers can be installed in the focus cabin at Parkes, and the choice of receivers is motivated both by technical availability and scheduling constraints); secondly, for an isolated transmitter on the surface of a rotating body we would expect intermittent behavior, which presents a potential opportunity for increasing the probability of interception with repeated observation; and finally, use of 0.7--4.0\,GHz Ultra-Wideband Low (`UWL') receiver (Hobbs et. al., in prep), is planned for future observations, which is the appropriate complement to observations with GBT. 

At Parkes, a total of \nhrpks\,hours of observations are used for this work, over the period 2016 November 16 to 2018 January 19 (MJD 57708--58137). During this time, \ncadencepks cadences covering a total of \nstarpks targets were observed: \nstarpksIsaacson from the 1702-star sample and \nstarpksExtra additional nearby stars (\reftab{tab:extra-stars}).

Work with other receivers, including the 21-cm multibeam \citep[MB, ][]{Staveley1996}, in addition to the already mentioned Ultra-Wideband Low \citep[UWL,][]{Manchester:2015}, is ongoing, and not included here.


\subsection{Data reduction pipeline}

A comprehensive overview of the BL data products and reduction pipelines is given by \citet{Lebofsky:2019}; here we provide a brief summary. Both GBT and Parkes use the same hardware and firmware to sample the incoming analog signals from the receiver, which we refer to as the signal processing `frontend'. This hardware, a 5\,Gsample/s, 8-bit digitizer and field-programmable gate array (FPGA) processing board, is provided by the Collaboration for Astronomy Signal Processing and Electronics Research \citep[CASPER, ][]{Hickish2016}. Detailed instrument descriptions are provided by  \citet{Macmahon:2017} and \citet{Price:2018}.

During observations, the frontend samples the dual-polarization receiver output at 8-bit resolution, then applies a polyphase filterbank to coarsely channelize the data into $\sim$2.92\,MHz bands, running firmware detailed by \citet{Prestage:2015}. The frontend FPGA boards output channelized data over 10Gb Ethernet to a cluster of high-performance compute nodes, each of which captures 187.5\,MHz of dual-polarization data. The compute nodes write 8-bit voltage-level products to disk in \texttt{raw} format\footnote{\texttt{raw}, \texttt{filterbank}, and \texttt{hdf5} formats are detailed by \citet{Lebofsky:2019}}. Each node is equipped with a Graphics Processing Unit (GPU), which is used to convert the voltage-level \texttt{raw} files into spectral data products, stored in \texttt{filterbank} format.

A total of three \texttt{filterbank} data products are generated: a high-spectral-resolution product with frequency and time resolution of $\sim$2.79\,Hz and $\sim$18.25\,s respectively, a mid-resolution product ($\sim$2.86\,kHz, $\sim$1.07\,s), and a high-time-resolution product ($\sim$366\,kHz, $\sim$349\,$\upmu$s). Here, as we are searching for the presence of narrowband signals, we analyse only the high-spectral-resolution product.

After observations are completed, the spectral products from each compute node are  combined into a single frequency-contiguous file, and converted into \texttt{hdf5} format. All data analyzed here are available online\footnote{\url{ https://breakthroughinitiatives.org/opendatasearch}}; final data volumes are 142\,TB and 77\,TB for Green Bank and Parkes, respectively.

\section{Methods} \label{sec:analysis}

\begin{figure*}
\begin{center}
\includegraphics[width=1.6\columnwidth]{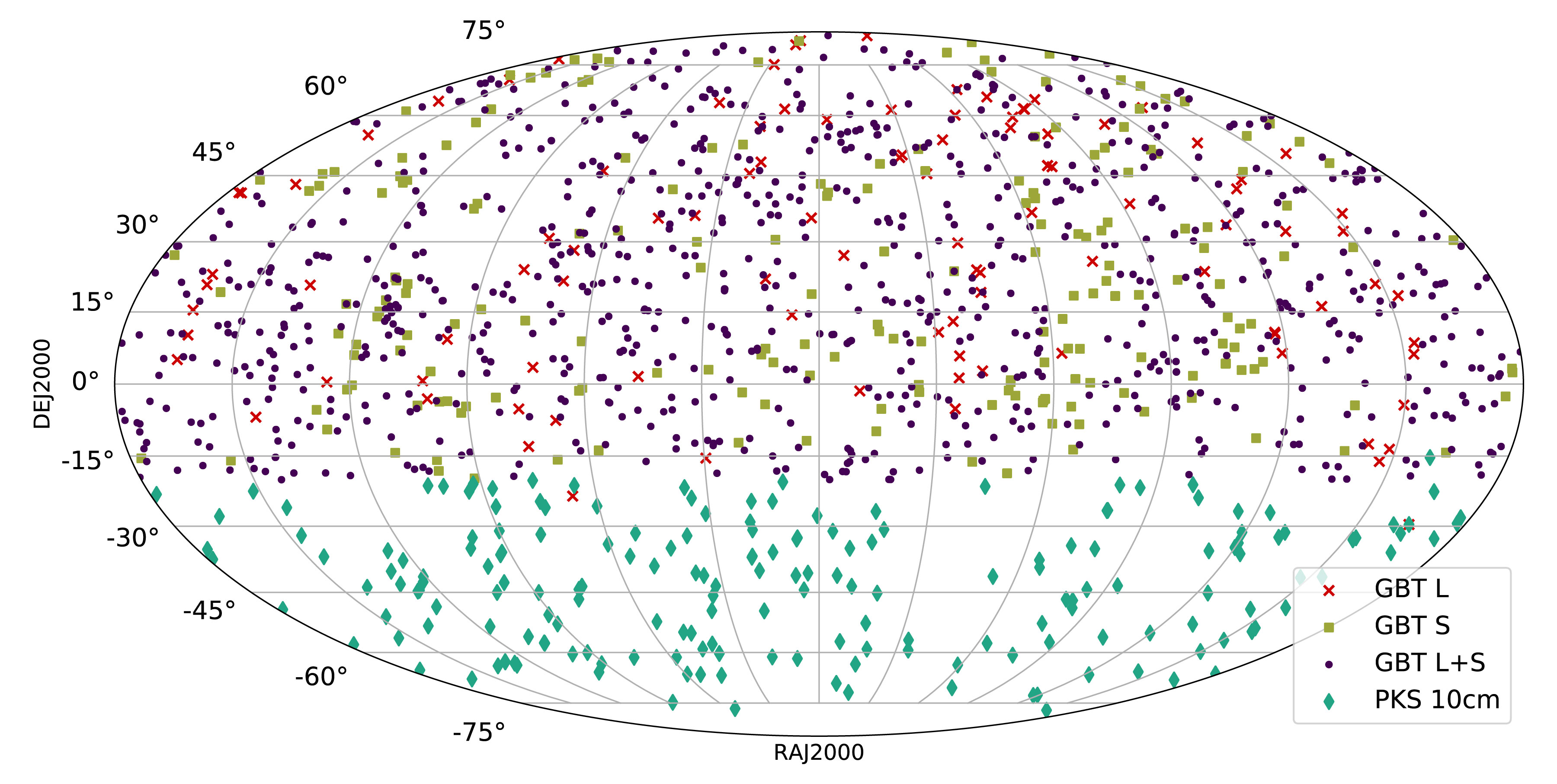}
\protect\caption{Distribution of observed sources in equatorial coordinates, taken from the 1702-star sample of \citet{Isaacson:2017}. Sources observed with Green Bank at both L-band and S-band are plotted in purple; sources only observed at L-band are plotted with red crosses; sources only observed at S-band are plotted with yellow squares; and sources observed with Parkes at 10cm are plotted with aqua diamonds.
\label{fig:allsky}
}
\end{center}
\end{figure*}

\begin{figure}
\begin{center}
\includegraphics[width=1.0\columnwidth]{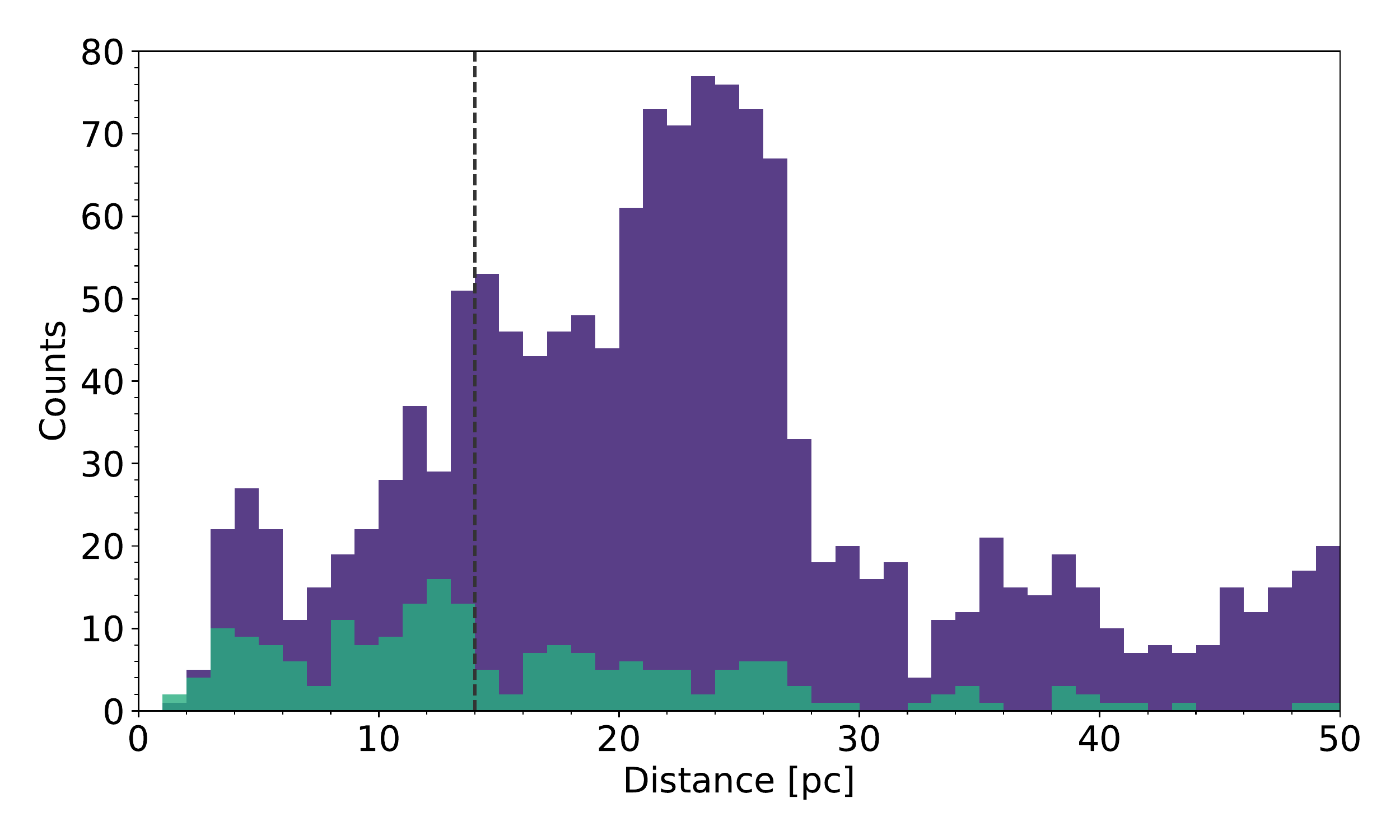}
\protect\caption{Histograms of the distances for sources shown in \reffig{fig:allsky}, observed with Green Bank (purple) and Parkes (aqua). 
\label{fig:distances}
}
\end{center}
\end{figure}


For a receiver fixed on Earth, any transmitter not also on the Earth's surface (or in geosynchronous orbit) will exhibit a time-dependent Doppler shift due to relative acceleration between the transmitter (ET) and receiver. 
The maximum Doppler shift in frequency, $\Delta\nu_{\rm{max}}$, depends upon the relative velocity $\Delta v$ of the transmitter, and transmitted frequency $\nu_{\rm{ET}}$:
\begin{equation}
    \Delta\nu_{\rm{max}} = \nu_{\rm{ET}}\left(\frac{\Delta v}{c}\right).
\end{equation}
Over short ($\sim$5\,minute) durations, the change in frequency
is well approximated as a linear function
\begin{equation}
    \nu(t) = \nu_0 + \dot{\nu}t,
\end{equation}
where $\nu_0$ is the frequency at $t=0$, and $\dot{\nu}$ is the shift in frequency (units Hz\,s$^{-1}$) due to Doppler motion, or drift rate. 
If after a time $t_{\rm{obs}}$ the product 
$\dot{\nu}\times t_{\rm{obs}}$
 is greater than the channel bandwidth $B$, 
signal power will be `smeared' across $N=\dot{\nu}\times t_{\rm{obs}} / B$ channels, lowering the detected signal to noise (S/N) in each channel by a factor $N^{1/2}$. This effect can be compensated for, if the observation is split into sub-integrations, by applying a shift to each sub-integration before integrating to form a final spectrum. Usually the sub-integration smearing is not corrected for---this approach is known as an incoherent search. If the drift rate is not known, a search across a range of trial drift rates can be conducted to identify the drift rate that optimizes detection; this search can be done by a brute-force approach or other means. Our detection algorithm \tseti uses a tree search algorithm which is optimized computationally \citep[][and references therein]{Enriquez:2017}.  

The maximum Doppler drift due to a body's rotation is given by
\begin{equation}
    \dot{\nu} = \frac{4 \pi^2 R}{ P^2}\frac{\nu_0}{c},
\end{equation}
where $c$ is the speed of light, $P$ is rotational (or orbital) period, and $R$ is the body (or orbit) radius. At the lower and upper frequency limits of our dataset (1.1--3.4\,GHz), Earth's daily rotation corresponds to drift rates of magnitude 0.12--0.38\,\Hzs; Earth's 1\,AU orbit imparts 0.02--0.06\,\Hzs. Here, we search drift rates between -4 to +4\,\Hzs, \diff{corresponding to fractional drift rates of 3.64\,nHz at 1.1\,GHz, and 1.15\,nHz at 3.45\,GHz. This rate allows for a wide range of planetary radii, spin periods, and orbital periods: \citet{Oliver:1971} advocate a 1\,nHz rate, based on that expected from an Earth-like planet with an 8-hr day. More recently, in simulations of planetary formation models, \citet{Miguel:2010} find a majority of planetary primordial rotation periods for $<10M_{\earth}$ 
 planets fall between 10 and 10000 hours; this sets a 0.65\,nHz maximum rate. Note that a putative transmitter that is not gravitationally bound would still exhibit a doppler drift imparted by Earth's rotation. }

\subsection{Dedoppler search analysis}

Following \citet{Enriquez:2017}, we use the \tseti incoherent dedoppler code \citep{Enriquez:2019}\footnote{\url{https://github.com/ucberkeleyseti/turbo_seti}} to search our high-resolution ($\sim$2.79\,Hz) Stokes-I data for drifting narrowband signals.

\tseti applies an efficient tree search algorithm, based on \citet{Taylor:1974} and \citet{2013ApJ...767...94S}. 
The tree algorithm efficiently computes the integrals over straight line paths. It is similar to the Hough transform \citep[e.g.][]{Leavers:1992}, which also computes integrals over straight line paths, but also applies an edge-detection step to convert input data into a binary image. The Hough transform itself can be thought of as a 2-dimensional discrete Radon transform \citep{Gotz:1995}. The tree dedispersion algorithm 
accelerates searches by reuse of redundant computations involved when searching similar slopes, 
which reduces the number of additions required to $n\rm{log}_{2}n$ from $n^2$, where $n$ is equal to both the number
of spectra and number of slopes searched.

By the radiometer equation, the noise within a coarse channel (without RFI present) follows a chi-squared distribution, as the digitized voltages are well approximated as zero-mean Gaussian random processes \citep{ThompsonMoranSwenson}. Here, we compute root-mean-square noise levels from the 90th central percentile of the power values to mitigate outliers in the distribution, due to the presence of narrowband features, and rolloff imparted by the shape of the polyphase filter.

The number of discrete frequency drift rates that are searched by \tseti
depends on $\dot{\nu}_{\rm{min}} = B/t_{\rm{obs}}$, roughly 0.01\,\Hzs\, for our high-frequency resolution data products. For our search to rates $\pm$4\,\Hzs, roughly 800 Doppler trials are performed.

We ran \tseti on all files, searching drift rates $\pm$4\,Hz, for narrowband signals with a S/N $\geq$10. We parallelized
processing tasks across nodes using a code called \tiddalik\footnote{\url{https://github.com/ucberkeleyseti/tiddalik}}, which distributes and executes tasks across nodes. As \tseti runs on a single file, processing is `pleasingly parallel' and can
be run without inter-task communication.

\tseti produces a list of `hits', i.e. detections above a given S/N, in a \texttt{.dat} plaintext file. We define a hit as the signal with highest S/N per
  channel over all the drift rates searched. Only the signal with the highest S/N within a window $\pm \dot{\nu}_{\rm{max}}\times t_{\rm{obs}} /2=\pm 600$\,Hz is recorded as a hit.
  
  We used the Python \textsc{Pandas}\footnote{\url{https://pandas.pydata.org}} package to read these files into a searchable table.  To quickly process multiple files, we used \textsc{Dask}\footnote{\url{https://dask.pydata.org}} to batch process multiple files in parallel; this is far less computationally intensive than \tseti and as such only a single compute node was required. To load data and read observational metadata from \texttt{filterbank} and \texttt{hdf5} formats, we use the \textsc{Blimpy}\footnote{\url{https://github.com/ucberkeleyseti/blimpy}} package \citep{Price:2019}.

\subsection{Data selection}\label{sec:data-sel}

After \tseti is run on each file, sets of files are grouped to form complete ABABAB or ABACAD cadences. Observation sets that are not part of a complete cadence are not analysed further in this work. We require that all files with a cadence contain 16 integrations (5 minutes). We then use the \texttt{find\_event} method of \tseti to search for hits that are present in all ON source observations, but not in OFF source observations. 

We refer to hits matching this criterion as an `event'. Specifically, any set of hits present in all ON observations in a frequency range calculated by $2 \dot{\nu_0} \times t_{\rm{obs}}$ and central frequency $\nu_n$ are selected. Where 
$\dot{\nu}_{\rm{min}} \leq   | \dot{\nu_0} | \leq  \dot{\nu}_{\rm{max}} $
is the drift rate of the hit in the first ON observation, and $\nu_n = \nu_0 + \dot{\nu_0} \times (t_n - t_0) $, while $t_0$ and $t_n$ are the observing start times of the first and $n$th observation respectively.  
Additionally, any set of hits for which there is at least one hit in the OFF observations within $\pm$600\,Hz of the hit frequency from the first ON observation would be discarded. This window corresponds to half the maximum searched drift of a signal over the period of the observation.

\subsection{Event grouping}\label{sec:event-grouping}

Discrete analysis of an event without regard to surrounding events does not provide a complete picture.  Events that are clustered in frequency, all of which exhibit the same drift rate, are likely to be associated with a single source of interference (or, indeed, technosignature).

We apply a simple grouping algorithm to assign events into groups, to aid in visualization and analysis. Events are grouped into frequency bins of width 125\,kHz, then in each bin the spacing between highest and lowest start frequency is computed, to compute an effective bandwidth $\Delta\nu_{\rm{event}}$, and central frequency $\nu_{\rm{event}}$. We refer to each cluster as an `event group'. 


\subsection{Event rejection and analysis}

We reject event groups where frequencies are within the GBT L-band or GBT S-band notch filters, but do not outright reject zero drift signals. Examples of events that pass all criteria are shown in Appendix \ref{sec:event-appendix}. Any event group that satisfies these filters is considered as a candidate signal and visually inspected. To do so, we plot the dynamic spectra of the events using the \textsc{Blimpy} package, for all on-source and reference pointings. We reject events when it is clear by eye that the event group is present in reference pointings, but was not detected above the \tseti S/N$>10$ threshold.

\section{Results} \label{sec:results}

We ran \tseti on all files with a complete observing cadence, finding a total of 51.71 million hits across the L-band, S-band, and 10-cm datasets (\reftab{tab:hit-counts}). Of these, 21,117 events were detected only in ON observations, which we clustered into 6,154 event groups. We treated analysis for each receiver separately, detailed below.

\begin{table}
\begin{center}
\caption{\label{tab:hit-counts} Summary of hits (signals above threshold), events (hits only in ON observations), event groups (clusters of events), and final events (groups with limited frequency extent).}

\begin{tabular}{lcccc}
\hline

Receiver & Hits & Events & Groups & Final \\
\hline
\hline
GBT L-band     & \nhitsgbtL  &  \neventgbtL   & \ngroupsgbtL  & \nfinalL \\
GBT S-band     & \nhitsgbtS  & \neventgbtS    & \ngroupsgbtS  & \nfinalS \\
Parkes 10-cm   & \nhitspks   & \neventpks     & \ngroupspks   & \nfinalpks   \\
\hline
\end{tabular}

\end{center}
\end{table}

\subsection{Green Bank}

\tseti was run on \ncadencegbtL L-band observing cadences (\nhrgbtL\,hr), and \ncadencegbtS S-band cadences (\nhrgbtS\,hr). We required that cadences had a full set of ON and OFF observations (ABACAD), that all subbands were present, and all had a cadence of 16 integrations, indicating that the observation was not interrupted. 

\subsubsection{L-band (1.10--1.90\,GHz)}
At L-band, a total of \nhitsgbtL hits were detected by \tseti above the S/N$>$10 threshold. The hit distribution as a function of drift rate and S/N are shown in \reffig{fig:gbt-hits-drift-L} and \reffig{fig:gbt-hits-snr-L}, respectively. Dashed lines at $\pm0.25$\,\Hzs\,correspond to maximum drift rate expected due to Earth's motion at the receiver's upper limit of 1.9\,GHz. A majority of hits (\nhitsgbtLzerodrift) are at zero drift, a large fraction (\nhitsgbtLnegativedrift) have negative drift rates, and a smaller fraction (\nhitsgbtLpositivedrift) have positive drift rates. The bias of signals toward negative drift rates is likely due to satellites in non-geosynchronous orbits, which accelerate with respect to the telescope \citep{Zhang:2006}. 

Hit density as a function of frequency is shown in \reffig{fig:gbt-L-freq}. Areas of high hit density are associated with known sources of RFI, in particular Global Positioning Systems satellites (GPS), operating at 1.155–-1.196\,GHz (L5) and 1.555–-1.596\,GHz (L1). Bands associated with GLONASS satellites at 1.592--1.619\,GHz (L1) and 1.192--1.212\,GHz (L3) also show high hit density, along with the satellite downlink band 1.525--1.560\,GHz.

Of the \nhitsgbtL hits, \neventgbtL events (hits found only in ON observations) were detected (\reffig{fig:gbt-hits-drift-L} and \reffig{fig:gbt-hits-snr-L}, light blue). This cut also excludes events within the GBT L-band notch filter (1.200–-1.341\,GHz). Events are concentrated at \diff{S/N $<$1000}, and at positive drift rates.  

The events are clustered into \ngroupsgbtL event groups, from \neventgbtLnstar\, unique stars. After visual inspection of these event groups, we do not find any signals that can not be attributed to RFI; see Appendix \ref{sec:event-appendix}.

\subsubsection{S-band (1.80--2.80\,GHz)}

\begin{figure*}
\centering 
\subfloat[Drift rate hit density, GBT L-band]{%
  \includegraphics[width=0.98\columnwidth]{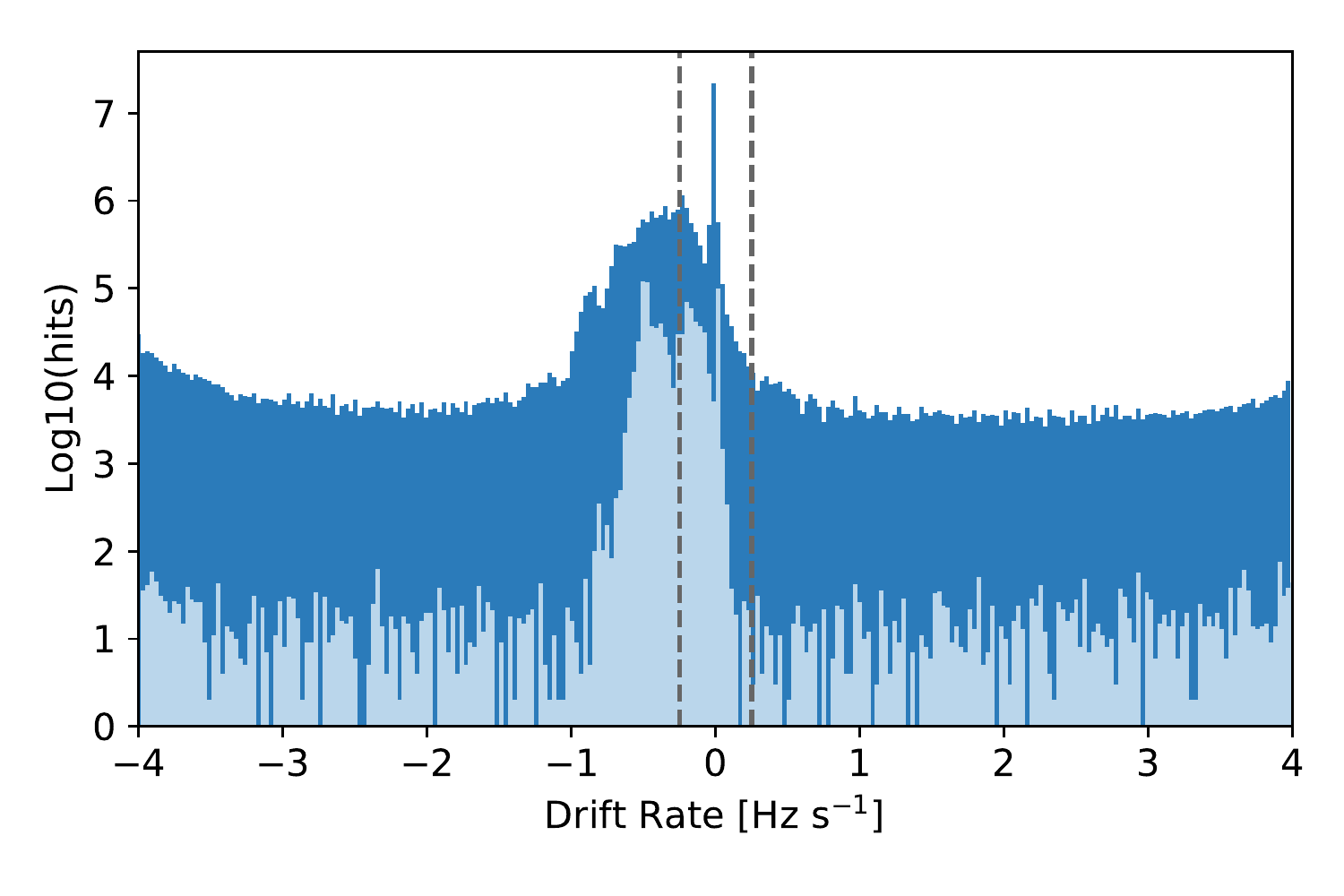}%
  \label{fig:gbt-hits-drift-L}%
}\qquad
\subfloat[Hit density as function of S/N, GBT L-band]{%
  \includegraphics[width=0.98\columnwidth]{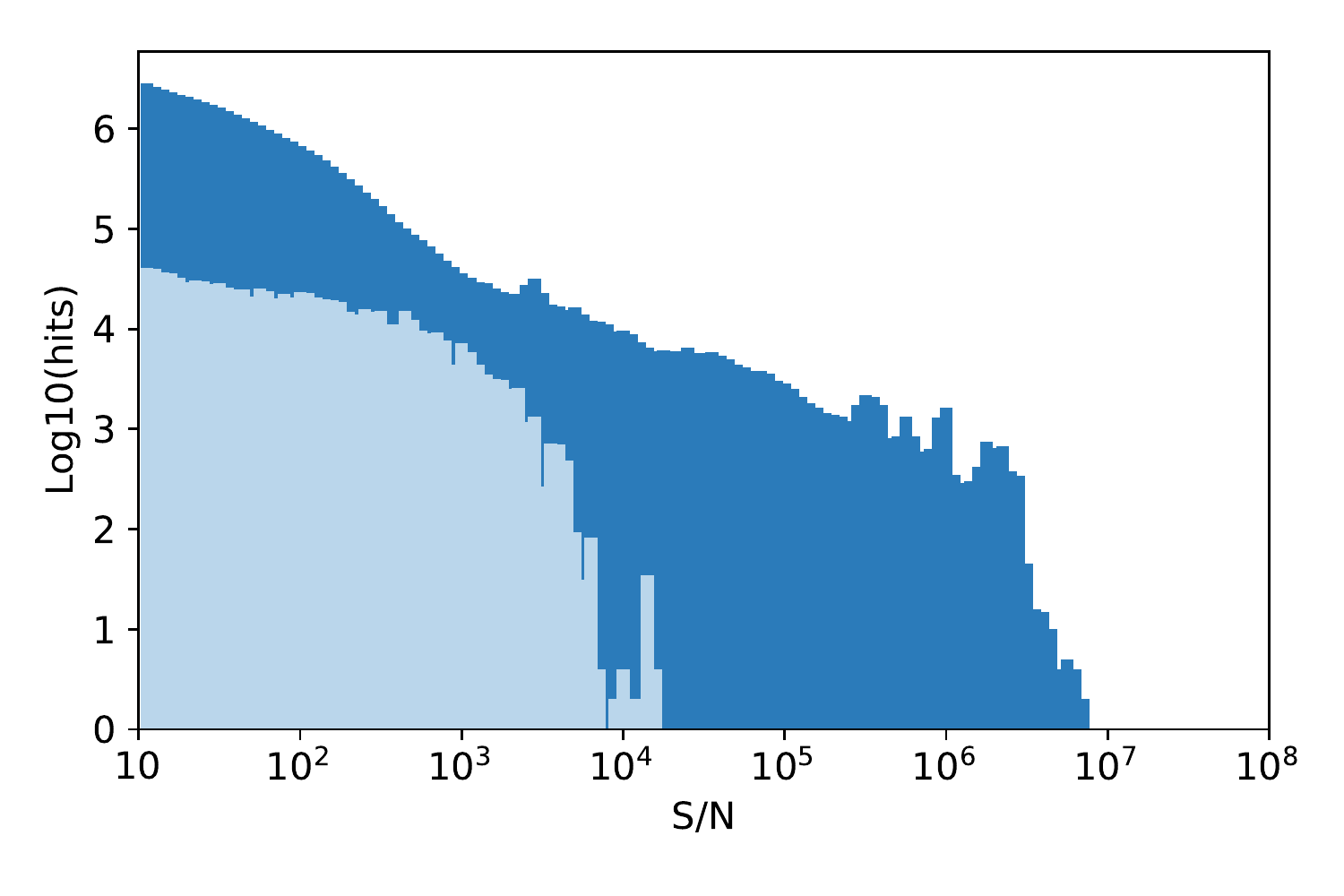}%
  \label{fig:gbt-hits-snr-L}%
}\qquad
\subfloat[Drift rate hit density, GBT S-band]{%
  \includegraphics[width=0.98\columnwidth]{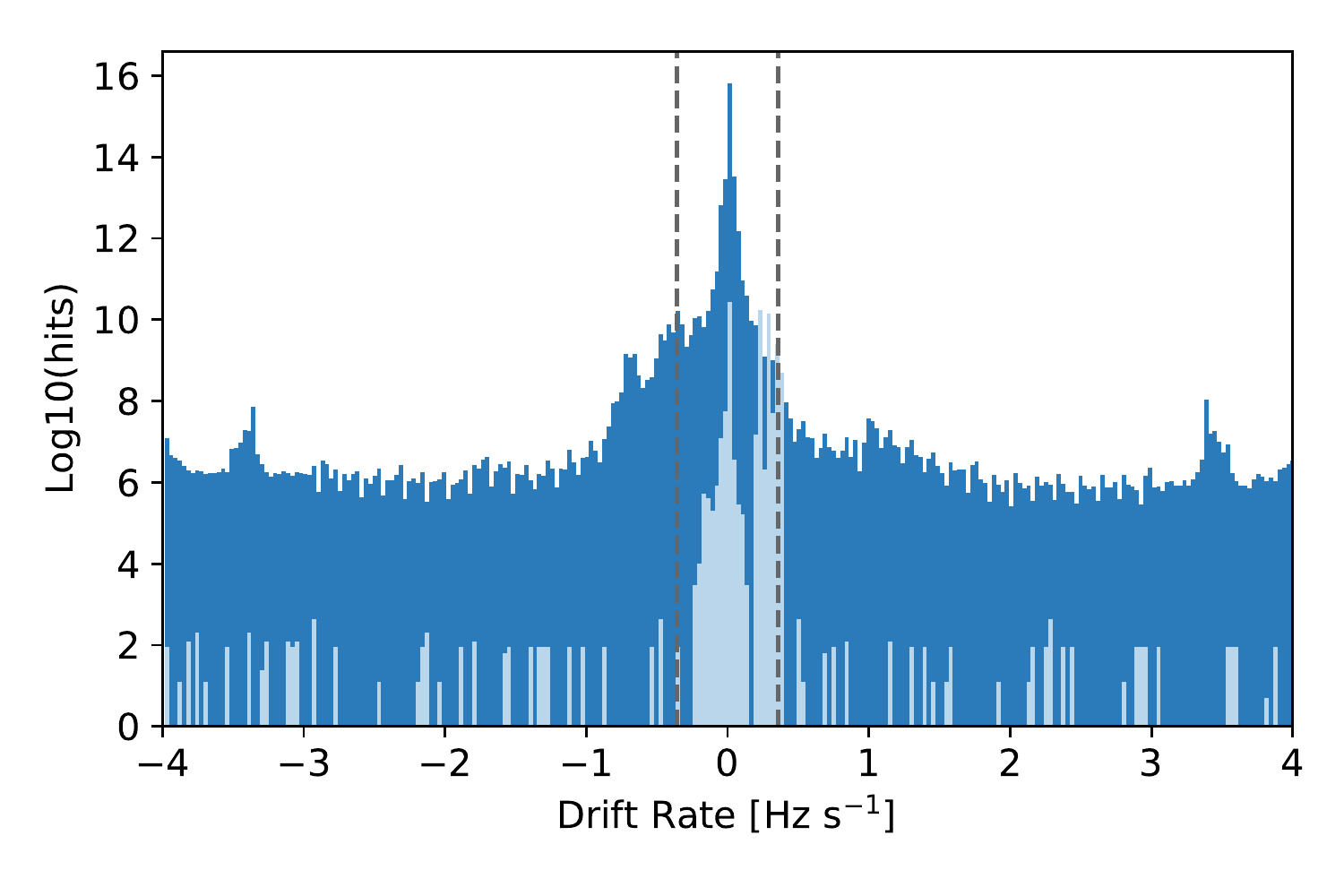}%
  \label{fig:gbt-hits-drift-S}%
}\qquad
\subfloat[Hit density as function of S/N, GBT S-band]{%
  \includegraphics[width=0.98\columnwidth]{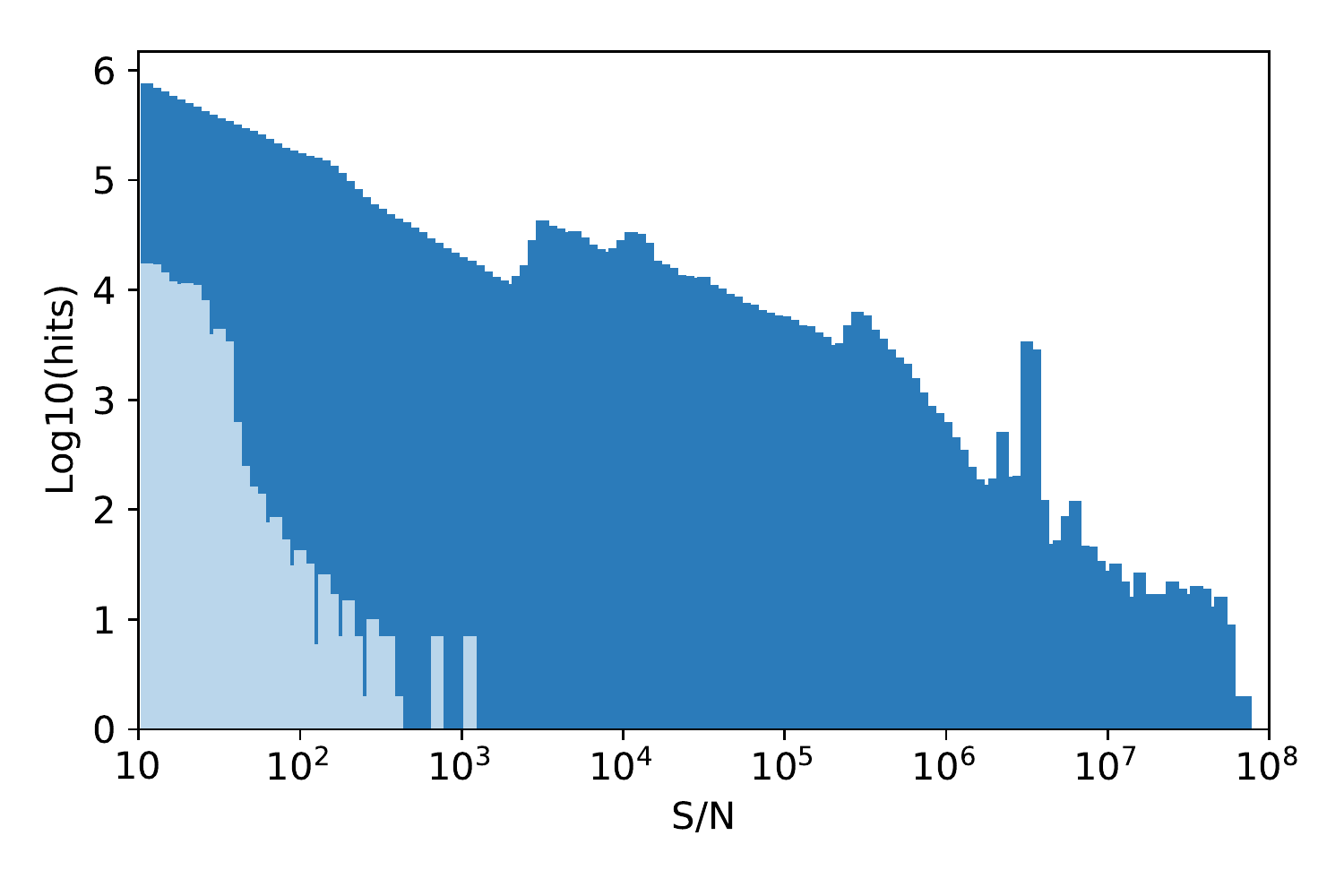}%
  \label{fig:gbt-hits-snr-S}%
}\qquad
\subfloat[Drift rate hit density, Parkes 10-cm]{%
  \includegraphics[width=0.98\columnwidth]{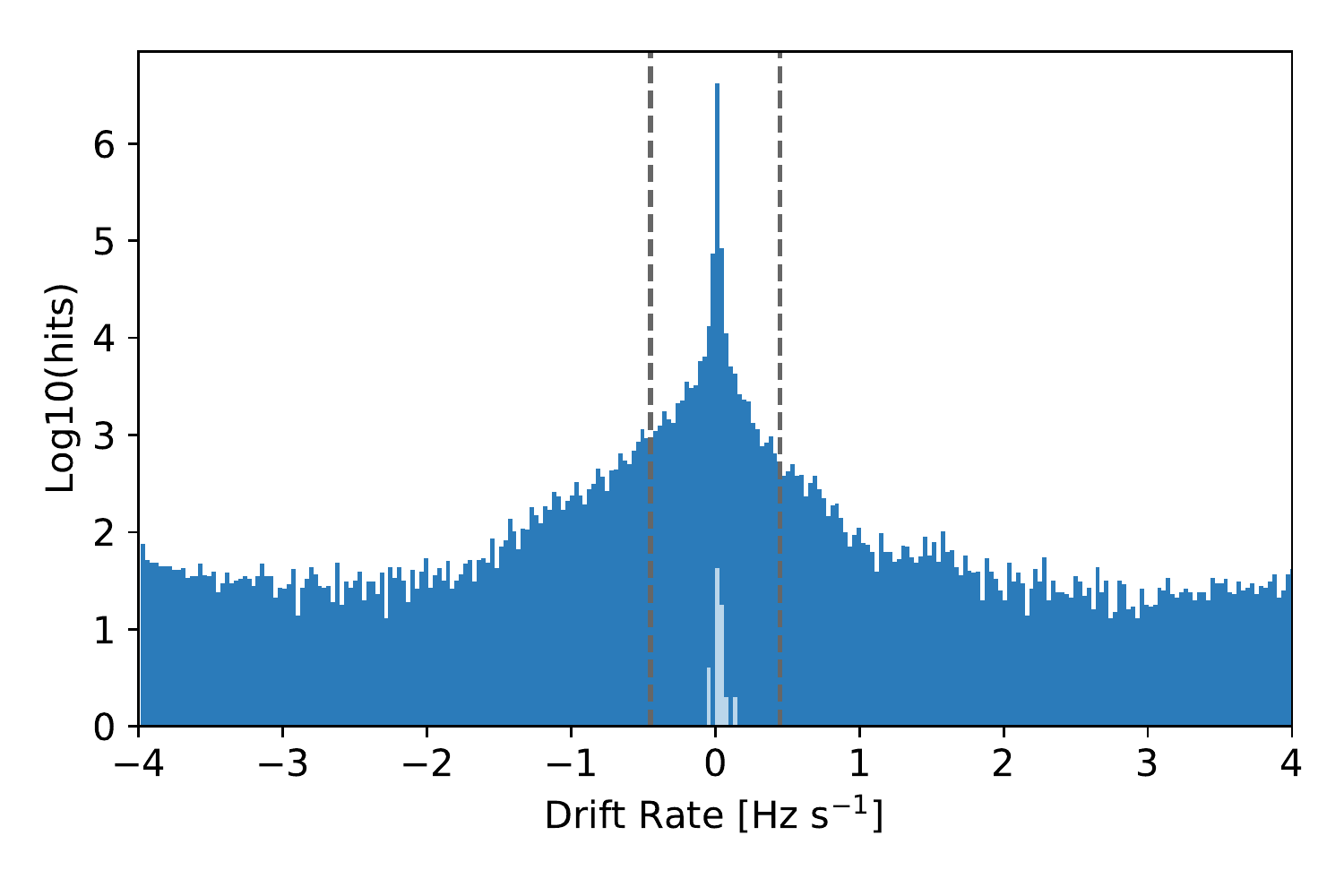}%
  \label{fig:pks-hits-drift}%
}\qquad
\subfloat[Hit density as function of S/N, Parkes 10-cm]{%
  \includegraphics[width=0.98\columnwidth]{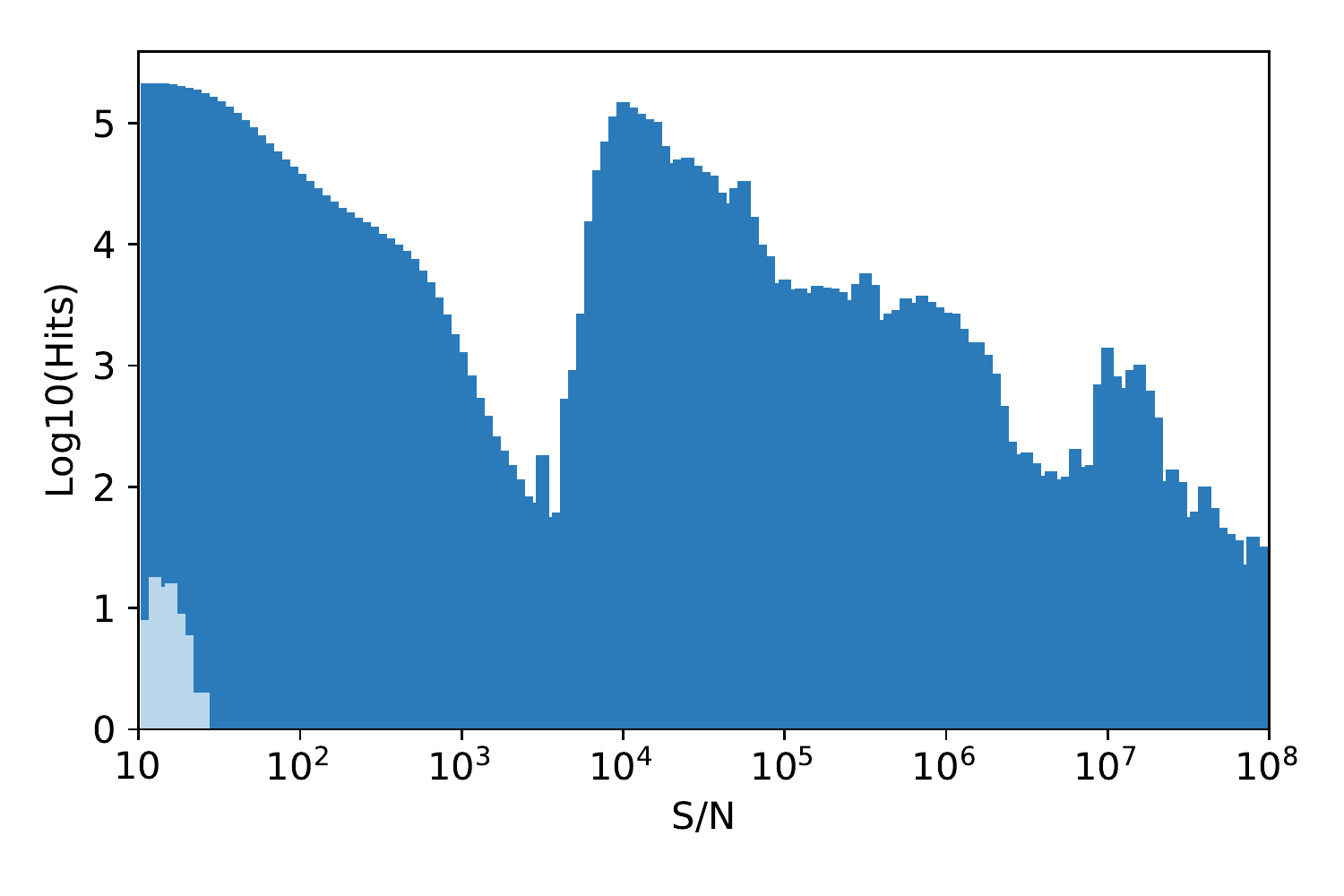}%
  \label{fig:pks-hits-snr}%
}\qquad
\caption{Histograms of hit (dark blue) and event (light blue) densities for GBT L-band, GBT S-band and Parkes 10-cm observations, for drift-rate (left) and as a function of signal to noise (right).\label{fig:histograms-all}}
\end{figure*}

At S-band, a total of \nhitsgbtS hits were detected; hit density as a function of drift rate and SNR are shown in \reffig{fig:gbt-hits-drift-S} and \reffig{fig:gbt-hits-snr-S}.  A majority of hits (\nhitsgbtSzerodrift\, out of 10.1M) are at zero drift; however a larger portion exhibit non-zero drift, with a small skew towards negative drift rates (\nhitsgbtSnegativedrift\, vs \nhitsgbtSpositivedrift). Dashed lines at $\pm0.36$ correspond to maximum drift rate expected due to Earth's motion at the receiver's upper limit of 2.7\,GHz. 

Of the \nhitsgbtS hits, \neventgbtS events (hits found only in ON observations) were detected (\reffig{fig:gbt-hits-drift-S} and \reffig{fig:gbt-hits-snr-S}, light blue). Events are concentrated at low S/N ($<10^3$), and at positive drift rates.  \diff{The peaks at $\pm 3.5$\,Hz\,s$^{-1}$ apparent in \reffig{fig:gbt-hits-drift-S} are associated with RFI around 1930.2\,MHz, within the cellular band, and as such may be due to cell phone activity near the observatory.} 

The hit density falls as S/N increases (\reffig{fig:gbt-hits-snr-S}), from $\sim$millions of hits per bin at S/N of 10, down to $\sim$tens of hits at S/N of $10^7$. Hit density as a function of frequency is shown in \reffig{fig:gbt-L-freq} and \reffig{fig:gbt-S-freq}. A large number of hits are attributable to RFI (see \ref{tab:fcc-allocs}).

After applying event grouping, a total of \neventgbtS event groups were identified. After visual inspection, we do not find any candidate signals not attributable to RFI (Appendix \ref{sec:event-appendix}).

\subsection{Parkes 10-cm (2.60--3.45\,GHz)}

We ran \tseti on the \ncadencepks observation cadences (\nhrpks\,hr); a total of \nhitspks\, hits were detected. Histograms showing hit density as a function of drift rate are shown in \reffig{fig:pks-hits-drift} and \reffig{fig:pks-hits-snr}. A majority of hits (4.16M out of \nhitspks) are at zero or within $\pm$0.015\,Hz (\reffig{fig:pks-hits-drift}). Outside of zero drift, there is a slight skew towards negative drift rates (134k vs 126k). Dashed lines at $\pm0.45$ correspond to maximum drift rate expected due to Earth's motion at the 3.45\,GHz upper limit of the receiver. 

The hit density over S/N (\reffig{fig:pks-hits-snr}) falls steadily until an S/N of $\sim$1000, after which it rises rapidly before falling again. \diff{This may indicate distinct populations of interferers with different characteristic signal strengths that are not isotropically distributed.}   

Hit density as a function of frequency is shown in \reffig{fig:pks-freq}. A large number of hits are associated with known RFI sources: 2.60--2.62\,GHz is 4G cellular service  downlink (band 7), and the 3.4\,GHz band is licensed to the Australian National Broadband Network (NBN).

Of the \nhitspks\, hits, only \ngroupspks event groups from 20 stars pass our selection criteria (\refsec{sec:data-sel}).  These events predominantly fall in the 3.40--3.45\,GHz band, and  are likely interference associated with NBN. Example events are shown in \reffig{fig:pks-example-events}. In all cases, the narrowband signal was detected with a S/N$>$10 in the ON source pointings, but not in OFF source pointings.  The effectiveness of the ON--OFF approach using Parkes at these frequencies indicates that this band is relatively quiet and, in regions where it isn't, at least relatively stable.

\section{Discussion}

\subsection{\diff{Comparison across receivers}}

\begin{table}
\begin{center}
\caption{\label{tab:rate-density} Rate densities for hits (signals above threshold), events (hits only in ON observations), and event groups (clusters of events).}

\begin{tabular}{lccc}
\hline

 &  L-band &  S-band    &  10-cm  \\
\hline
\hline
Hit rate density & \multirow{2}{*}{111.1}  & \multirow{2}{*}{20.0}   & \multirow{2}{*}{10.8}  \\
($\times10^{-6}$\,hits\,Hz$^{-1}$hr$^{-1}$)   \\
Event rate density  & \multirow{2}{*}{47.9}  & \multirow{2}{*}{10.1}    & \multirow{2}{*}{0.18}  \\
($\times10^{-9}$\,events\,Hz$^{-1}$hr$^{-1}$) \\
Grouped event rate density & \multirow{2}{*}{13.5}   & \multirow{2}{*}{3.1}    & \multirow{2}{*}{0.1}   \\
($\times10^{-9}$\,groups\,Hz$^{-1}$hr$^{-1}$)  \\
\hline
\end{tabular}

\end{center}
\end{table}

\diff{A summary of hits across the L-band, S-band and 10-cm datasets is shown in \reftab{tab:hit-counts}. Broadly, the number of hits decreases with increasing receiver frequency. Taking into account receiver bandwidth and observation time, we can compare `hit rate density', that is, number of hits per unit bandwidth per hour. This, and similarly defined event rate densities, are shown in \reftab{tab:rate-density}.  }

\diff{A higher hit rate density corresponds to higher levels of RFI occupancy. However, hit rate is dependent on the S/N threshold, the sensitivity of the telescope, and observation strategy: a less sensitive telescope will report fewer hits above a given S/N for the same RFI environment; similarly, observations toward the horizon or known RFI sources will have higher rate densities. As such, direct comparison between observatories and observing campaigns is nuanced.}

\diff{Nevertheless, the rate densities in \reftab{tab:rate-density} do inform us about broad RFI trends. The L-band receiver has a hit rate density more than 5$\times$ higher than that of the S-band, and 10$\times$ that of the Parkes 10-cm receiver. Lower rate densities with the S-band receiver (for the same observational approach) indicate the RFI environment is cleaner over 1.9--2.8\,GHz than 1.1--1.9\,GHz at the GBT. The grouped event rate density is 135$\times$ higher with the GBT L-band than the Parkes 10-cm, showing that a larger fraction of events pass our RFI identification at L-band than at 10-cm. We attribute this to a lack of satellites, or other rapidly moving sources, that transmit over 2.6--3.4\,GHz.}

\diff{Plots of hit density as a function of frequency are given in \reffig{fig:freqs-all}. Areas of high hit density are shaded;  corresponding federal allocations are given in \reftab{tab:fcc-allocs}. In the US, the Federal Communications Commission (FCC\footnote{\url{http://fcc.gov}}) and National Telecommunications and Information Administration (NTIA\footnote{\url{http://ntia.gov}}) oversee spectrum allocations; in Australia this is done by the Australian Communications and Media Authority (ACMA\footnote{\url{http://acma.gov.au}}). These agencies coordinate with the International Telecommunications Union (ITU\footnote{\url{https://www.itu.int}}), who also regulate space-based frequency allocations.
}

\begin{table}
\begin{center}
\caption{\label{tab:fcc-allocs} Radio frequency spectrum allocations, 
for bands with high hit densities (see \reffig{fig:freqs-all}).
}

\begin{tabular}{ll}
\hline

 Band & Federal Allocation \\
 (MHz) &  \\
\hline
\hline

1164--1215 & Global Navigation Satellite System (GNSS)$^\dagger$ \\
1350--1390 & Air traffic control (ATC)$^\dagger$ \\
1525--1535 & Mobile-satellite service (MSS)$^\dagger$ \\
1535--1559 & MSS$^\dagger$ \\
1559--1610 & GNSS$^\dagger$ \\
1675--1695 & Geostationary operational environmental \\
           & satellite (GOES)$^\dagger$  \\
1850--2000 & Personal communications services (PCS)$^\dagger$ \\
2025--2035 & GOES $^\dagger$  \\
2100--2120 & NASA Deep Space Network use (DSN)$^\dagger$ \\
2180--2200 & MSS$^\dagger$ \\
2200--2290 & Earth exploration satellite (EES)$^\dagger$ \\
2620--2690 & Mobile Telecommunications (MT)$^\ddagger$ \\
3425--3492.5 & National Broadband Network (NBN)$^\ddagger$  \\
\hline
\multicolumn{2}{l}{$^\dagger$FCC/NTIA Table of Frequency Allocations} \\
\multicolumn{2}{l}{$^\ddagger$ACMA Australian Radiofrequency Spectrum Plan} \\
\end{tabular}

\end{center}
\end{table}

\subsection{Combined project metrics and figures of merit}

The search space for SETI signals is vast.  \citet{Tarter:2001} describe the search space as a `nine-dimensional haystack'; this metaphor is continued in \citet{Wright:2018}, who detail a method to compute a `haystack fraction' that quantifies how complete a SETI search is. The haystack fraction is but one of several figures of merit (FoM) that can be used as heuristics to compare SETI searches. 

An historical FoM is the Drake Figure-of-Merit \citep[DFM;][]{Drake:1984}, which is given by:
\begin{equation}
	\rm{DFM}=\frac{\Delta \nu_{\rm{tot}} \Omega}{\textit{F}_{\rm{min}}^{3/2}} ,
\end{equation} 
where $\Delta \nu_{\rm{tot}}$ is the observing bandwidth, $\Omega$ is the sky coverage, and $F_{\rm{min}}$ is the minimum detectable flux in W/m$^2$. The -3/2 index on $F_{\rm{min}}$ encompasses distance-to-volume scaling ($d^3$), and sensitivity scaling ($d^{-2}$).

As we use three receivers with varying fields-of-view, system temperatures, and bandwidths, we compute a combined DFM$_{\rm{tot}}$:
\begin{equation}
	\rm{DFM}_{\rm{tot}}=\sum_i^N{DFM_i},
\end{equation} 
i.e. the sum of DFMs; larger DFM values are better. The 
$\rm{DFM}_{\rm{tot}}$ for this project is $9.2\times$ than that of \cite{Enriquez:2017}. The combined sky coverage for all the observations was 22.1 squared degrees, in contrast with 10.6 squared degrees presented by \citet{Enriquez:2017}. 

We note that the above formulation of the DFM assumes a common channel bandwidth. For a narrowband signal, $F_{\rm{min}}$ depends upon the signal-to-noise threshold S/N$_{\rm{min}}$, and may be calculated as
\begin{equation}
	F_{\rm{min}} = {\rm{S/N_{min}}} \frac{2 k_B T_{\rm{sys}}}{A_{\rm{eff}}}
	\sqrt{\frac{B}{n_{\rm{pol}} t_{\rm{obs}} }}, \label{eq:S_min}
\end{equation} 
where $k_B$ is the Boltzmann constant, $T_{\rm{sys}}$ is the system temperature, $A_{\rm{eff}}$ is the effective collecting area of the telescope, $B$ is the channel bandwidth, and $n_{\rm{pol}}$ is the number of polarizations. Note that $F_{\rm{min}}$ (flux) is related to flux density $F_{\rm{min}}=S_{\rm{min}}/\delta\nu_{t}$, where $\delta\nu_{t}$ is the bandwidth of the transmitting signal. We have chosen a value of unity in this work. 

\citet{Wright:2018} presents a formalism in which one defines `boundaries' to specify an N-dimensional survey space, or `haystack'. One can then compute what fraction of a given haystack a survey probes. Using their boundaries, \citet{Wright:2018} compute a haystack fraction of $3.8\times10^{-19}$ for the observations presented in \citet{Enriquez:2017}. For our L-band, S-band, and 10-cm observations, we compute haystack fractions $1.23\times10^{-18}$, $7.44\times10^{-19}$, and $3.37\times10^{-19}$. These correspond to 3.24$\times$, 1.96$\times$, and 0.88$\times$ times that of \citet{Enriquez:2017}, respectively. The integrated value taking all the observations together is 6.08$\times$.

The DFM and haystack fraction are useful heuristics when comparing surveys. However, neither the DFM nor the haystack fraction take into account the distance to survey targets: they treat observations of nearby stars equal to a patch of seemingly blank sky. For this reason, \citet{Enriquez:2017} define the `Continuous Waveform Transmitter Rate Figure of Merit', or TFM:
\begin{equation}
	{\rm{TFM}}=\eta\frac{\rm{EIRP}_{\rm{min}}}{N_{\rm{star}}} \frac{\nu_c}{\Delta \nu_{\rm{tot}}}, 
\end{equation} 
where $\nu_{c}$ is the central observing frequency, $N_{\rm{star}}$ is the number of stars observed, and $\rm{EIRP}_{\rm{min}}$ is the minimum detectable equivalent isotropic radiated power (EIRP, in W), and $\eta$ is a normalisation factor. The value  $(\nu_{\rm{tot}} / \nu_{c}) / N_{\rm{star}} $ encompasses fractional bandwidth and number of sources, and is referred to as the transmitter rate. The $\rm{EIRP}_{\rm{min}}$ for a given target increases with the distance squared:
\begin{equation}
	{\rm{EIRP_{min}}}=4 \pi d^2 F_{\rm{min}}.  
\end{equation} 
The EIRP$_{\rm{min}}$ for our GBT observations is $2.1\times10^{12}$ W, and $9.1\times10^{12}$ W for Parkes observations at the 50 pc maximum distance of the \citet{Isaacson:2017}) sample. 

\begin{figure*}
\centering 
\includegraphics[width=1.8\columnwidth]{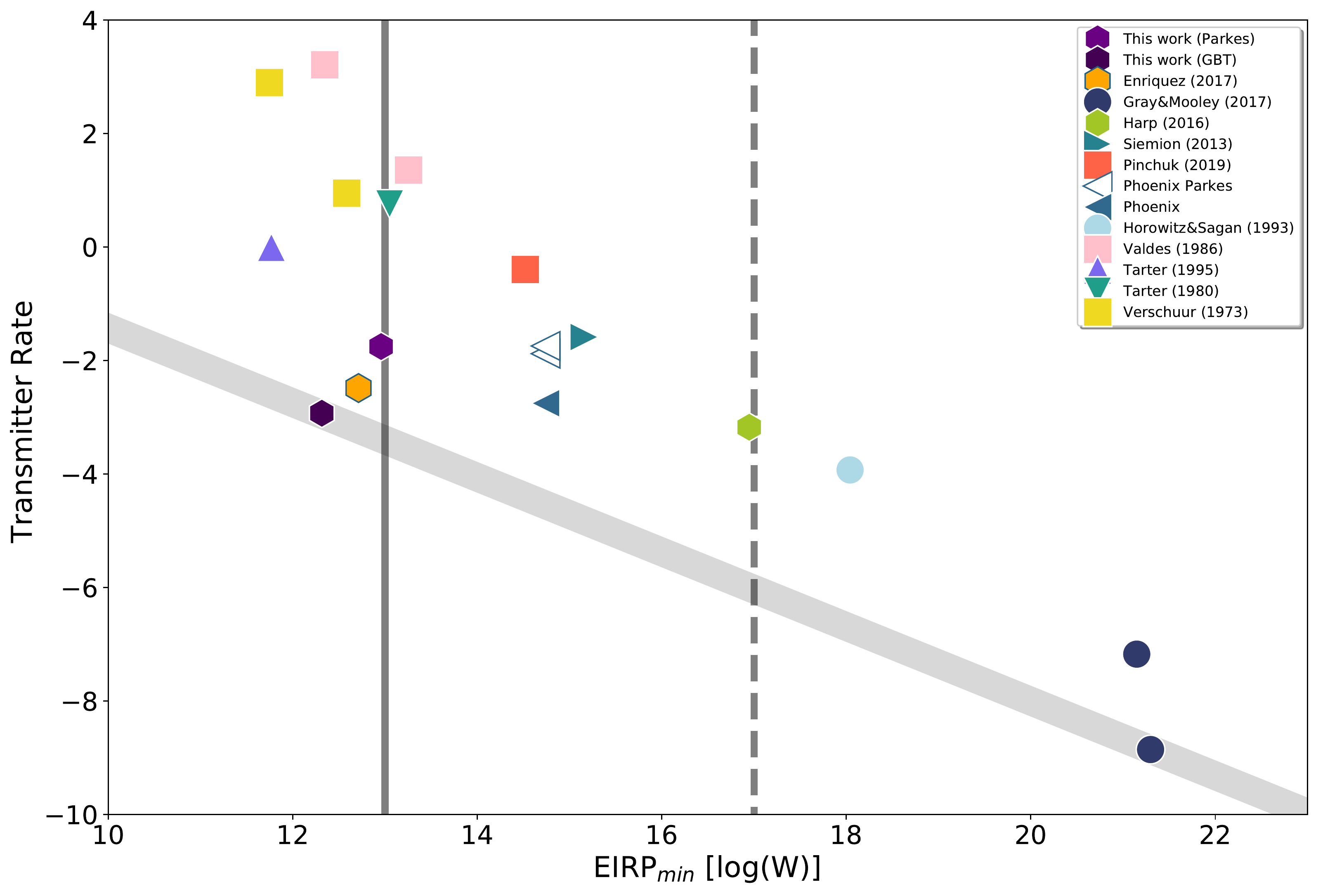}%
\caption{Transmitter Rate comparison with other historical projects.
The transmitter rate, $( N_{\rm{star}}(\frac{\nu_{c}}{\nu_{\rm{tot}}}))^{-1}$, is plotted on logarithmic axes against the minimum detectable power, EIRP$_{\rm{min}}$, based on the distance to the farthest star in the sample. Points toward the bottom of this plot represent surveys with large numbers of star targets and high fractional bandwidth; points toward the left represent surveys where the product of sensitivity and distance to targets is lower. The solid and dashed vertical lines represent the EIRP of the Arecibo planetary radar, and total power from the Sun incident on Earth, respectively. \diff{The diagonal grey line is a fit between the most constraining data points for transmitter rate and EIRP$_{\rm{min}}$.}
\label{fig:TFM}}
\end{figure*}

Numerically lower TFM scores represent more sensitive and more complete surveys. For comparison, the TFM for this work is 6.95$\times$ smaller than for \citet{Enriquez:2017}; comparisons against other surveys are shown in \reffig{fig:TFM}.

\subsection{Limits on narrowband technosignatures}

We find no evidence for narrowband transmitters from observations of our target stars above the EIRP$_{\rm{min}}$ values of $2.1\times10^{12}$\,W for GBT observations and $9.1\times10^{12}$\,W for Parkes observations. 

It is difficult to place limits on the existence of putative transmitters in the direction of the star targets, due to the presence of RFI, potential intermittency/periodicity of the transmission, or that our data analysis is insensitive to a given signal due to pipeline limitations (see \refsec{sec:limitations}). Nevertheless, one may still place a probabilistic upper limit on the prevalence of putative continuous narrowband transmitters above  EIRP$_{\rm{min}}$, assuming that such transmitters are rare. That is, one may compute a conditional probability of detecting a signal, should it exist above EIRP$_{\rm{min}}$ and within the observing band, by treating each star target as a trial within a Poissonian distribution. 

We make a conservative estimate that a given observation has a probability of $P=0.5$---to account for potential RFI obscuration---of detecting a narrowband signal at a random frequency within the observed band (above the EIRP$_{\rm{min}}$). For GBT L-band, 882 star targets were observed; treating these as discrete trials we place a limit with 95\% confidence that fewer than 0.45\% of stars have narrowband transmitters above EIRP$_{\rm{min}}$ of $2.1\times10^{12}$\,W. For S-band, this limit is 0.37\% of stars (EIRP$_{\rm{min}}$ $2.1\times10^{12}$\,W), and 2.0\% based on the Parkes 10-cm data (EIRP$_{\rm{min}}$ $9.1\times10^{12}$\,W). 

These limits are coarse, and could be improved by more careful consideration of several aspects. Firstly, one could inject signals at various drift rates into real observational data, to compute signal recovery statistics. Secondly, one could run Monte Carlo simulations in which transmitters are placed at different distances, and their signal properties are drawn from varied probability distributions, following a Bayesian approach \citep[e.g.][]{Grimaldi:2018}; this is an avenue for future investigation.

\begin{figure*}
\centering 
\subfloat[GBT L-band: hits density per MHz]{%
  \includegraphics[width=2.0\columnwidth]{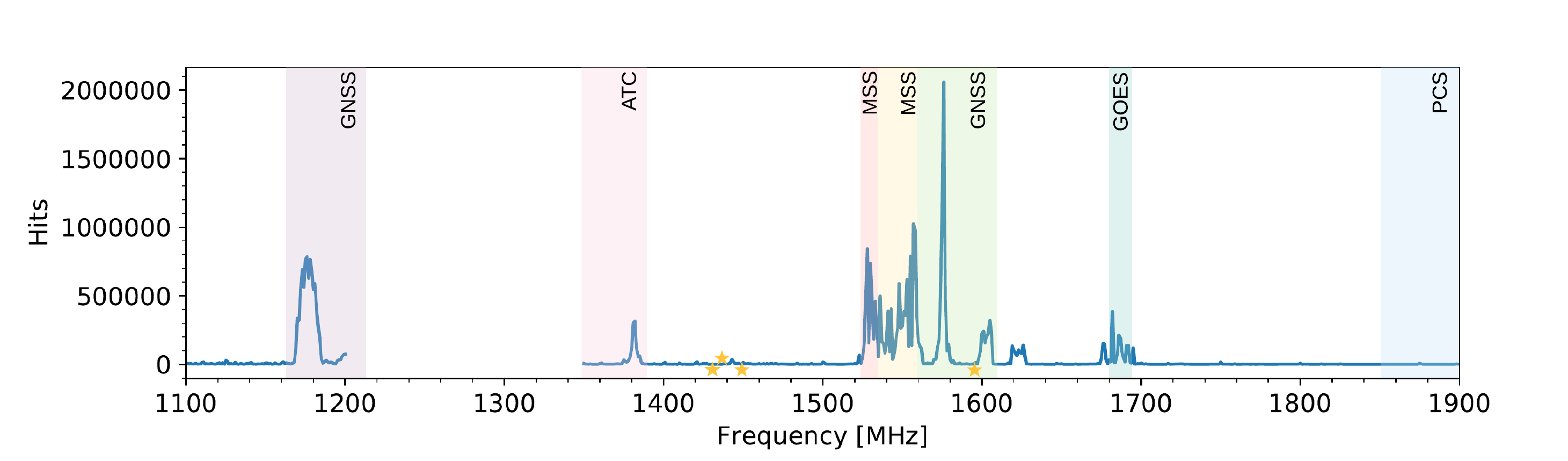}%
  \label{fig:gbt-L-freq}%
}\qquad
\subfloat[GBT S-band: hits density per MHz]{%
  \includegraphics[width=2.0\columnwidth]{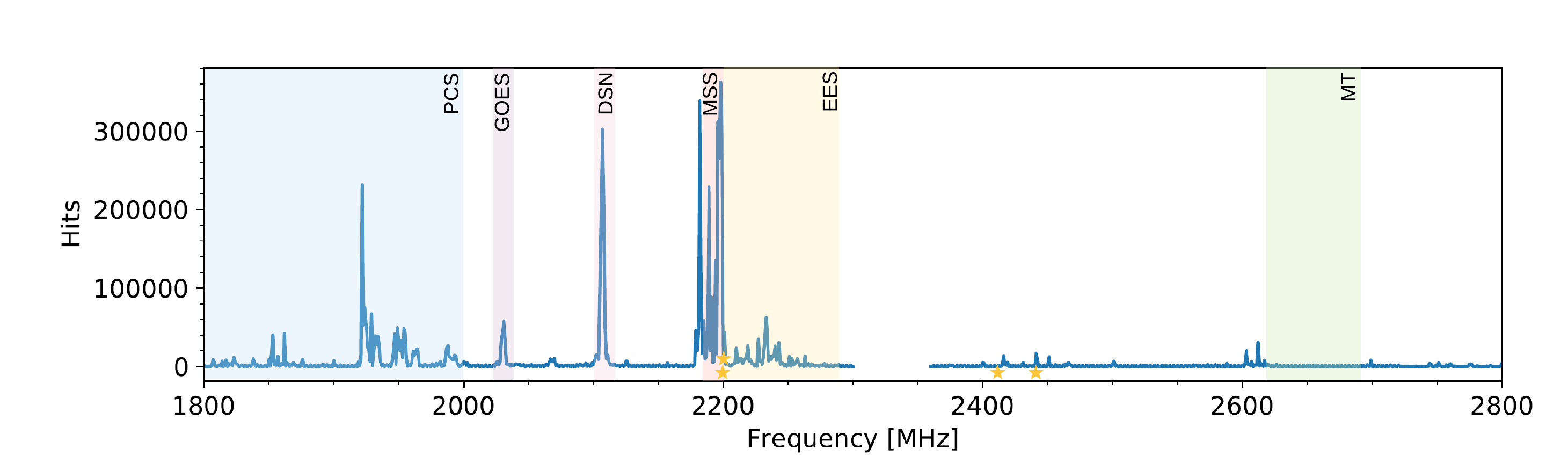}%
  \label{fig:gbt-S-freq}%
}\qquad
\subfloat[Parkes 10-cm: hits density per MHz]{%
  \includegraphics[width=2.0\columnwidth]{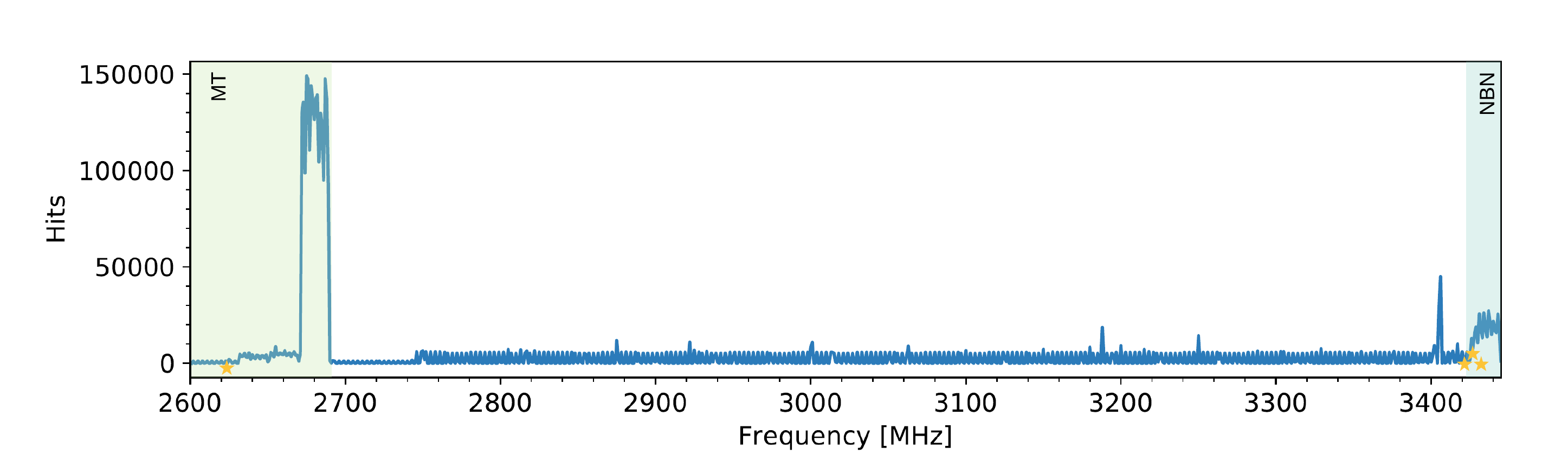}%
  \label{fig:pks-freq}%
}
\caption{Frequency distribution of hits produced by the \tseti pipeline, over the bands of the GBT L-band, GBT S-band, and Parkes 10-cm receivers. Frequencies with large hit densities (shaded) are associated with known sources of RFI; see \reftab{tab:fcc-allocs}. Events of interest as detailed in Appendix\,\ref{sec:event-appendix} \label{fig:freqs-all} are marked with stars.}
\end{figure*}

\subsection{Comparison to Enriquez et. al. (2017)}

In addition to new observations, we reanalysed the observations reported by \citet{Enriquez:2017} over 1.10--1.90\,GHz, using a lower S/N cutoff (10 vs. 25). The result is that our sensitivity is better by a factor of 2.5$\times$, but as a side effect, our false positive rate also increases. A number of events were recorded where a signal was present in the OFF observation, but was not detected above the S/N threshold; \citet{Enriquez:2017} avoided this by requiring a higher a S/N (25) for ON observations than OFF observations (20). 
We also reanalysed data across a broader range of drift rates, expanding from $\pm2$\,\Hzs~ to $\pm4$\,\Hzs. A side effect of the larger drift rate is that the window used to avoid redundant detections, given by $\pm\dot{\nu}_{\rm{max}}\times t_{\rm{obs}}/2$, is larger. 

\subsection{Comparison to previous 1.80--3.45\,GHz searches}
SETI searches over the combined 1.80--3.45\,GHz range of the GBT S-band and Parkes 10-cm receiver have been conducted previously, but to a lesser extent than the so-called `water-hole' between 1.42--1.67\,GHz. As part of the SERENDIP-II survey, \citet{Werthimer:1986} observed a 32\,kHz band around 2.25\,GHz using a 210\,ft antenna. Project Sentinel \citep{Horowitz:1985} and project META \citep{Horowitz:1993} ran narrowband searches around 2.84\,GHz (twice the 21-cm line frequency). \citet{Tarter:1995} observed 24 Solar-type stars using a system with 8\,MHz bandwidth at 8-cm and 12-cm wavelengths as part of the High Resolution Microwave Survey; this survey was defunded by congress before completion. Project Phoenix also covered this band, but published details on the observations are sparse \citep{Backus:2004}. To our knowledge, no archival data from any of these projects is publicly available.

More recently, the Allen Telescope Array \citep{Harp:2016} observed 9293 stars sporadically over 1--9\,GHz (averaging 785 MHz of observed band per star), and report a minimum detectable flux density $S_{\rm{min}}$=271\,Jy at 3\,GHz. This is roughly 7$\times$ more primary targets, but at 17--39$\times$ lower sensitivity, with a lower range of drift rates searched. By Eq.\,\ref{eq:S_min}, to reach an equivalent sensitivity would require observations between 289 to 1521 times longer. \citet{Harp:2016} searched drift rates of $\pm 2$\,\Hzs, over the 1.1--3.4\,GHz band.

\subsection{Comparison to other recent GBT searches}

Recent SETI searches were also undertaken at the GBT by \citet{Margot:2018} and in follow-up work by \citet{Pinchuk:2019} (henceforth M\&P). In both cases, the GBT L-band receiver was used, but different data analysis approaches were applied. Careful comparison of the two approaches to identify their relative advantages is invaluable. Here, we discuss the differences and similarities between these campaigns and BL L-band observations, to identify areas where future analyses can be improved.

\subsubsection{Observational Strategy}

Both \mgo and BL employed a similar observational strategy whereby targets were observed multiple times. \mgo observed each target twice for 150\,s per pointing, whereas we observed each target three times, for 300\,s per pointing. Our target selection, detailed by \citet{Isaacson:2017}, draws from a morphologically-diverse selection of stars, containing most types of stars existing within 50 pc; \mgo selected targets with known exoplanets, predominantly GKM type stars from the \emph{Kepler} field, as well as two nearby planet-hosting M-dwarfs. Combined, a total of 26 targets were observed in the \mgo sample, over 130\,minutes. 

\begin{figure}
    \centering
    \includegraphics{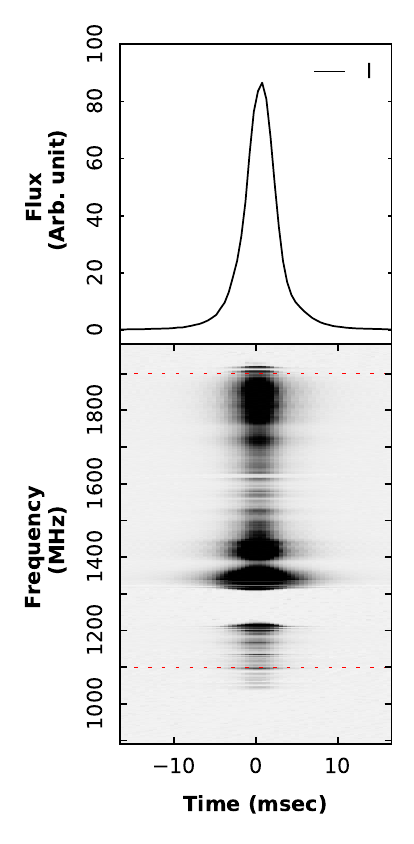}
    \caption{Observation of PSR J0826+2637 at L-band from the GBT using the BL backend. The top panel shows the integrated profile while the bottom panel shows dynamic spectra. The notch filter is clearly evident at frequencies between 1.2--1.33\,GHz. The red dotted-lines show the edges of the frequency band between 1.1--1.9\,GHz. A very clear pulsar detection can be seen well across these limits. }
    \label{fig:pulsar}
\end{figure}

\citet{Margot:2018} and \citet{Pinchuk:2019} only analyzed data within the nominal 1.15--1.73\,GHz passband of the receiver. Apart from elevated system temperature due to loss in aperture efficiency, we find no impediment to use of the full 1.10--1.90\,GHz band, although we note that both BL and \mgo avail themselves of the 1.20--1.34\,GHz notch filter to suppress nearby air surveillance radar (see Figure~\ref{fig:pulsar}).

{}

\subsubsection{2-bit requantization}
The most significant differences arise at the data recording level. \mgo employs the older GUPPI processing system, which records data as 2-bit quantized voltages \citep{2013ApJ...767...94S, Prestage:2015}. In contrast, BL records data at 8-bit resolution. BL then converts the recorded voltages into spectral products, resulting in a $\sim$50$\times$ reduction in data volume. While BL can archive 8-bit voltages, only a subset of voltages are retained due to storage limitations. Also, while voltage-level data products are more flexible than (fixed-resolution) spectral products, storing 2-bit data would require 12.5$\times$ increase in storage capacity (or 12.5$\times$ decrease in observing time). 

Requantizing to lower bit depth has several negative effects. The first is that the dynamic range---the ratio between largest and smallest possible values---is limited. The dynamic range in decibels available within $N$ bits is\footnote{\url{https://www.analog.com/media/en/training-seminars/tutorials/MT-229.pdf}}:
\begin{equation}
    \rm{DR = 20\,log_{10}(2^N-1) \,\,\,\, [dB]}
\end{equation}
For 2 and 8-bit data, the dynamic range afforded is 9.54\,dB and 48.13\,dB, respectively. 
Any signal that saturates the available dynamic range will be distorted. Impulsive RFI can introduce harmonic distortions, and interfere with nominal requantization. During requantization, the system bandpass must generally be subtracted dynamically, using scaling factors that change over small ($\sim$second) time windows. If the scaling factors are not preserved, bandpass information---a useful diagnostic---is lost. 

Quantization efficiency---the relative loss in S/N due to quantization--- drops from 0.99912 for 8-bits down to 0.881154 for 2-bit data, assuming optimal level settings \citep{Thompson:2007}. From a SETI perspective, the end result of requantization to 2 bits is that S/N threshold (e.g. S/N$>$10) would need to be lowered (S/N$>$8.8) to retrieve the same number of hits. The limited dynamic range also places a limit on the maximum S/N achievable, and strong hits will exhibit harmonic distortions that may register as extraneous hits. For these reasons, in situations where we store voltage-level data, we retain an 8-bit resolution.

\subsubsection{Drift rates and S/N threshold}

\citet{Margot:2018} use a window of size $\pm\dot{\nu}_{\rm{max}}\times t_{\rm{obs}}\approx \pm1500$\,Hz to remove redundant detections. Following \citet{Enriquez:2017}, we use $\pm\dot{\nu}_{\rm{max}}\times t_{\rm{obs}}/2$, i.e. half this value, but we fix the window size to $\pm600$\,Hz when comparing ON and OFF observations for RFI rejection. An issue with such maximum-drift-based approaches is that as the drift rate searched increases, the fraction of `blanked' band also increases, so candidate signals may be discarded, and metrics such as the DFM may be overestimated. To combat this, \citet{Pinchuk:2019} instead require that detections do not cross in time-frequency space. For continuity with \citet{Enriquez:2017}, we do not implement such a strategy here.  

\citet{Pinchuk:2019} estimate that the \citet{Margot:2018} DFM was overestimated by $\sim$5\% due to blanking. We compute average `blanked' fractions of 0.9\%, 0.2\% and 0.1\% for GBT L, GBT S and Parkes 10-cm observations, excluding notch filters and a 1\,kHz region around each hit, so this effect is negligible. Signal rejection filters will also affect metrics such as the DFM; we emphasize that figures of merit should be treated as heuristics for comparison of observational campaigns only.

\subsubsection{Event grouping and rejection}

\citet{Pinchuk:2019} grouped hits into $\sim$kHz bins, and discarded all hits in bins with high hit density. This is similar to our event grouping approach, however we promote events for visual inspection. Both of these approaches could likely be improved by identifying other signal properties (e.g. bandwidth, kurtosis), and forming a larger-dimensional parameter space in which to cluster signals. 

\subsection{Pipeline limitiations\label{sec:limitations}}
 
Based on the analysis of events, we identify several limitations of our current pipeline and corresponding areas of improvement. Firstly, it is often clear by visual analysis that a signal is indeed present in an OFF observation, but was below the S/N threshold required. The false-positive rate can be decreased by setting variable thresholds for OFF source pointings that account for the fact that sources well off-axis to the observing direction can nevertheless have varying apparent power in our ON and OFF source positions. Another possible method is to compute the cross-correlation between ON and OFF pointings over a set of lags and search directly for signals present common in ON and OFF at a lower total threshold.

Due to how the S/N is calculated, the S/N for events with bandwidth greater than a single channel is underestimated. By decimating in frequency (i.e. averaging over steps of $2^N$ channels), the S/N for wider bandwidth signals will increase until the signal is no longer resolved in frequency (e.g. as employed by \citet{2013ApJ...767...94S}). This approach is already used commonly in RFI flagging codes \citep[e.g.][]{Offringa:2012}. The estimation of S/N is also sensitive to the estimate of noise levels: in areas of high RFI occupancy, noise level estimates will be affected by the presence of RFI. 

For frequency-resolved signals, the S/N can also be improved by averaging across the bandwidth of the signal. One could use a hierarchical frequency decimation approach, searching optimal drift rates ranges at each stage to ensure the drift rate does not exceed $B/t_{\rm{obs}}$, which leads to smearing across channels. \diff{In a recent paper, \citet{sheikh:2019} advocates a rate as high as 200\,nHz, so as to be sensitive to a larger class of bodies, including exoplanets with highly eccentric orbits (e.g. HD 80606b) and small semimajor axes such as Kepler-78 b.}

Our frequency and drift rate grouping algorithm is simplistic, and could be improved using methods from machine learning (ML), such as $k$-means clustering. Drift rate and frequency are only two signal properties that could be used for grouping events, and with proper labelling, grouping could also take into account bandwidth, signal kurtosis or other assessments of modulation type. With appropriate training, ML methods can also be used to self-identify features \citep{Zhang:2019}.

\section{Conclusions}

As part of the BL program, we searched \nstar\,nearby stars taken from the \citet{Isaacson:2017} sample for technosignatures, using data from the Green Bank and Parkes telescopes. We used three receivers, spanning a combined range of 1.10--3.45\,GHz, and found no compelling candidates that are not attributable to radio interference. Our \tseti pipeline searched for narrowband signals exhibiting time-variable frequency drift due to Doppler acceleration, finding over 51 million hits above our S/N threshold. Of these hits, we identified 6154 event groups that passed our automated verification tests; however, none of these passed closer manual inspection and cross-referencing against known RFI. 

Combined, these observations constitute the most comprehensive survey for radio evidence of advanced life around nearby stars ever undertaken, improving on the results of \citet{Enriquez:2017} in both sensitivity and number of stars.  Together with other recent work from the resurgent SETI community, we are beginning to put rigorous and clearly defined limits on the behavior of advanced life in the universe.  We note that significant additional observational and theoretical work remains to be done before we are able to make general statements about the prevalence of technologically capable species.

With respect to the specific search described here, our analysis is currently confined to only spectrally narrow drifting signals using our highest resolution data product. A high-time-resolution data product will be searched for pulsed signals in future work, and a refined drifting spectral line search will be undertaken covering wider bandwidths. Further, BL observations with the GBT and Parkes are ongoing, with the GBT C-band (3.9--8.0\,GHz), GBT X-band (8.0--11.6\,GHz), and Parkes UWL (0.7--4.0\,GHz) receivers. Observations of the Galactic Plane are also being undertaken, using the Parkes multibeam receiver (1.2--1.6\,GHz). Observations are also planned with the MeerKAT telescope, and other partner facilities such as the Murchison Widefield Array.

\section{Acknowledgements}

Breakthrough Listen is managed by the Breakthrough Initiatives, sponsored by the
Breakthrough Prize Foundation. The Parkes radio telescope is part of the Australia Telescope National Facility which is funded by the Australian Government for operation as a National Facility managed by CSIRO. The Green Bank Observatory is a facility of the National Science Foundation, operated under cooperative agreement by Associated Universities, Inc. We thank the staff at Parkes and Green Bank observatories for their operational support.

\software{
\textsc{Blimpy} \citep{Price:2019}, 
\tseti \citep{Enriquez:2019},
\textsc{Astropy} \citep{price2018astropy},
\textsc{H5py} \citep{collette_python_hdf5_2014},
\textsc{Dask} \citep{dask:2016},
\textsc{Pandas} \citep{mckinney2010data},
\textsc{Matplotlib} \citep{hunter2007matplotlib},
\textsc{Numpy} \citep{oliphant2006guide},
\textsc{Scipy} \citep{scipy:2001},
\textsc{Jupyter} \citep{Kluyver:2016aa}
}

\bibliographystyle{aasjournal}
\bibliography{references}


\appendix

\section{Example events}\label{sec:event-appendix}

\begin{figure*}
\centering 
\subfloat[HIP54677]{%
  \includegraphics[width=0.45\columnwidth]{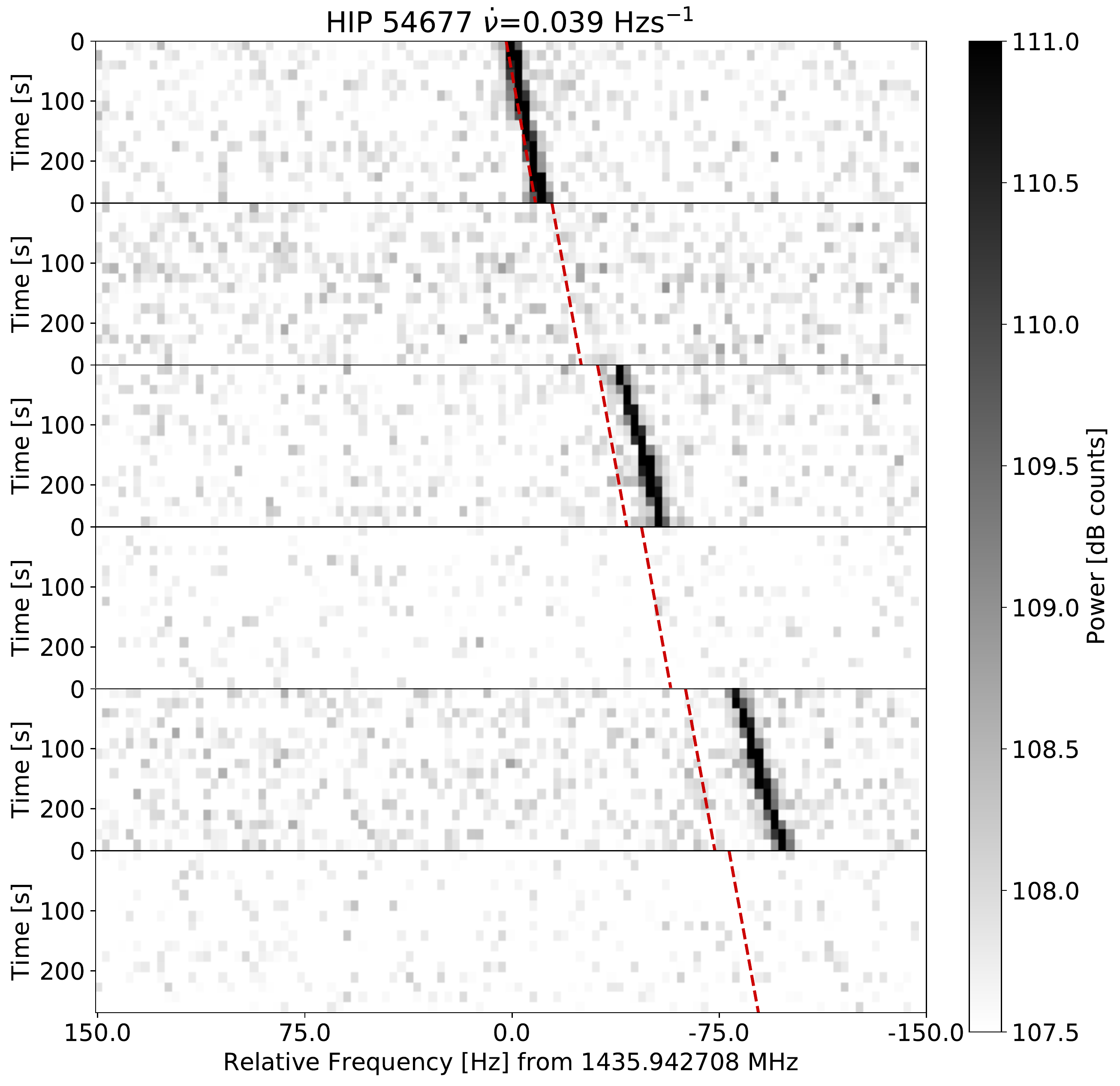}%
  \label{fig:L-event1}%
}\qquad
\subfloat[HIP54677]{%
  \includegraphics[width=0.45\columnwidth]{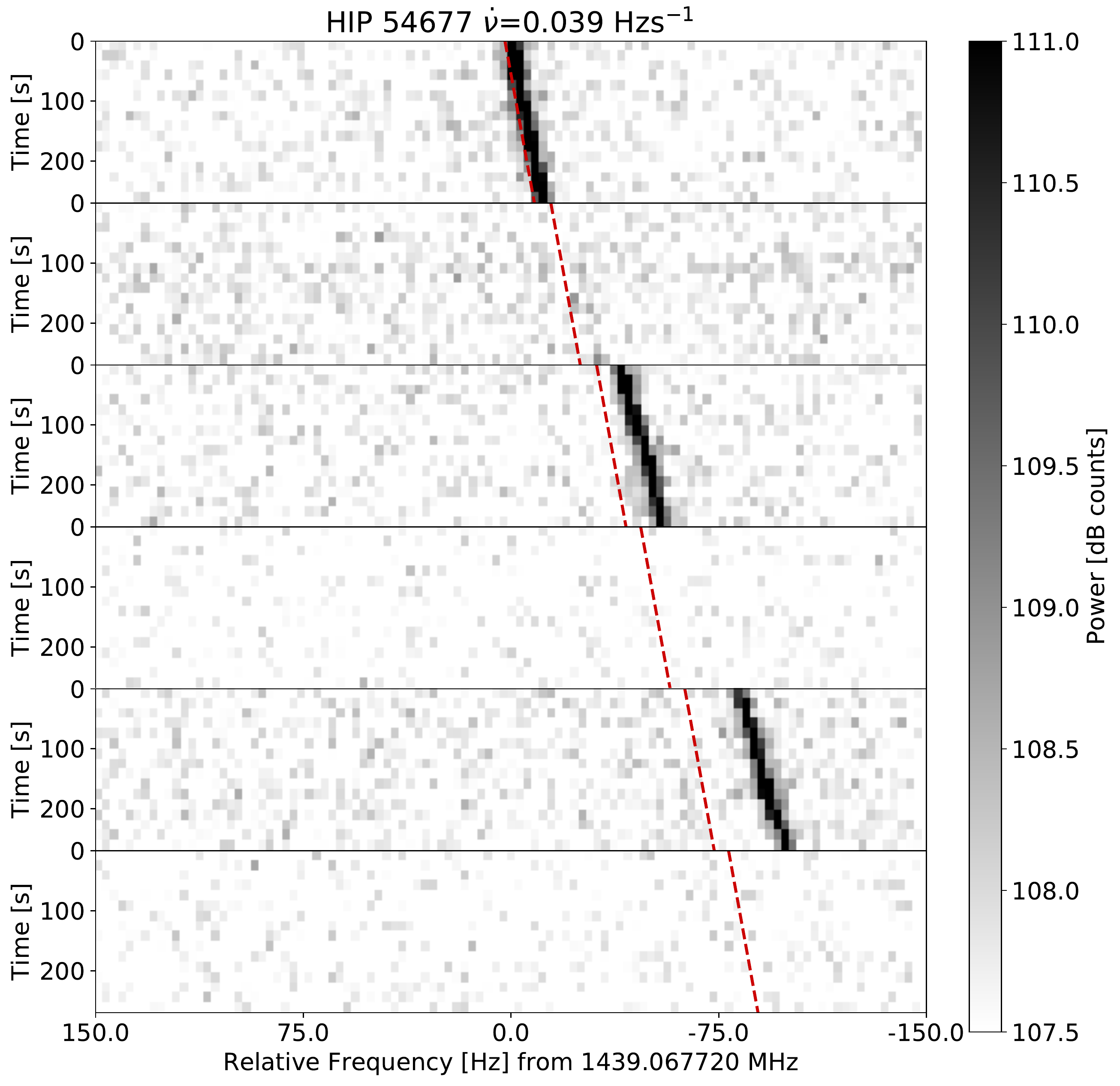}%
  \label{fig:L-event2}%
}

\subfloat[HIP10064]{%
  \includegraphics[width=0.45\columnwidth]{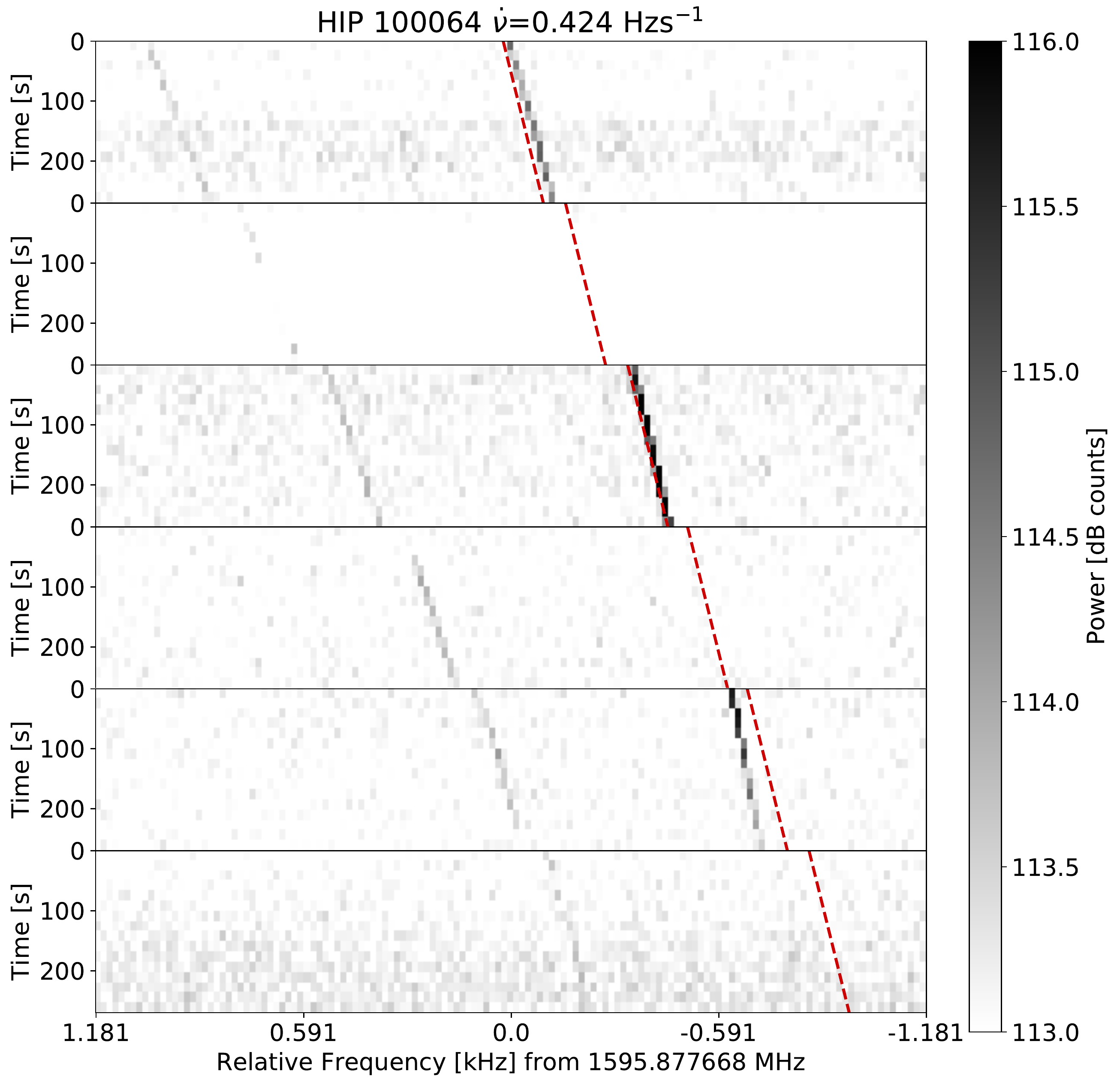}%
  \label{fig:L-event3}%
}\qquad
\subfloat[HIP1444]{%
  \includegraphics[width=0.45\columnwidth]{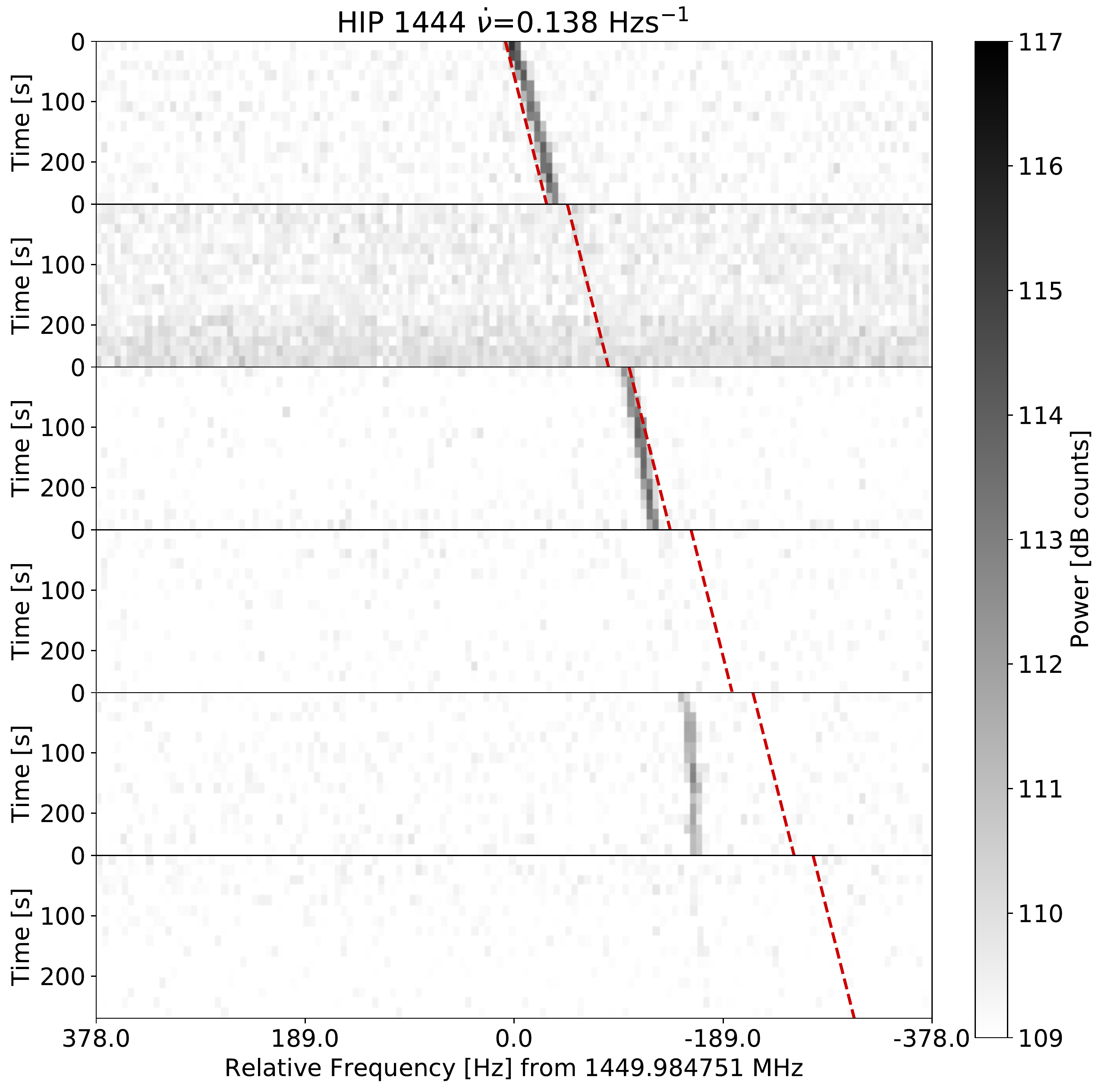}%
  \label{fig:L-event4}%
}
\caption{\label{fig:L-example-events}
Dynamic spectrum for selected \tseti events from GBT L-band observations. Each subfigure shows a full ABACAD cadence; each of the six panels represents ON and OFF source, consecutively. The red dashed lines
show the drift rate as detected by the pipeline; a small frequency offset has been added for visualization.}
\end{figure*}

\begin{figure*}
\centering 
\subfloat[HIP91699]{%
  \includegraphics[width=0.45\columnwidth]{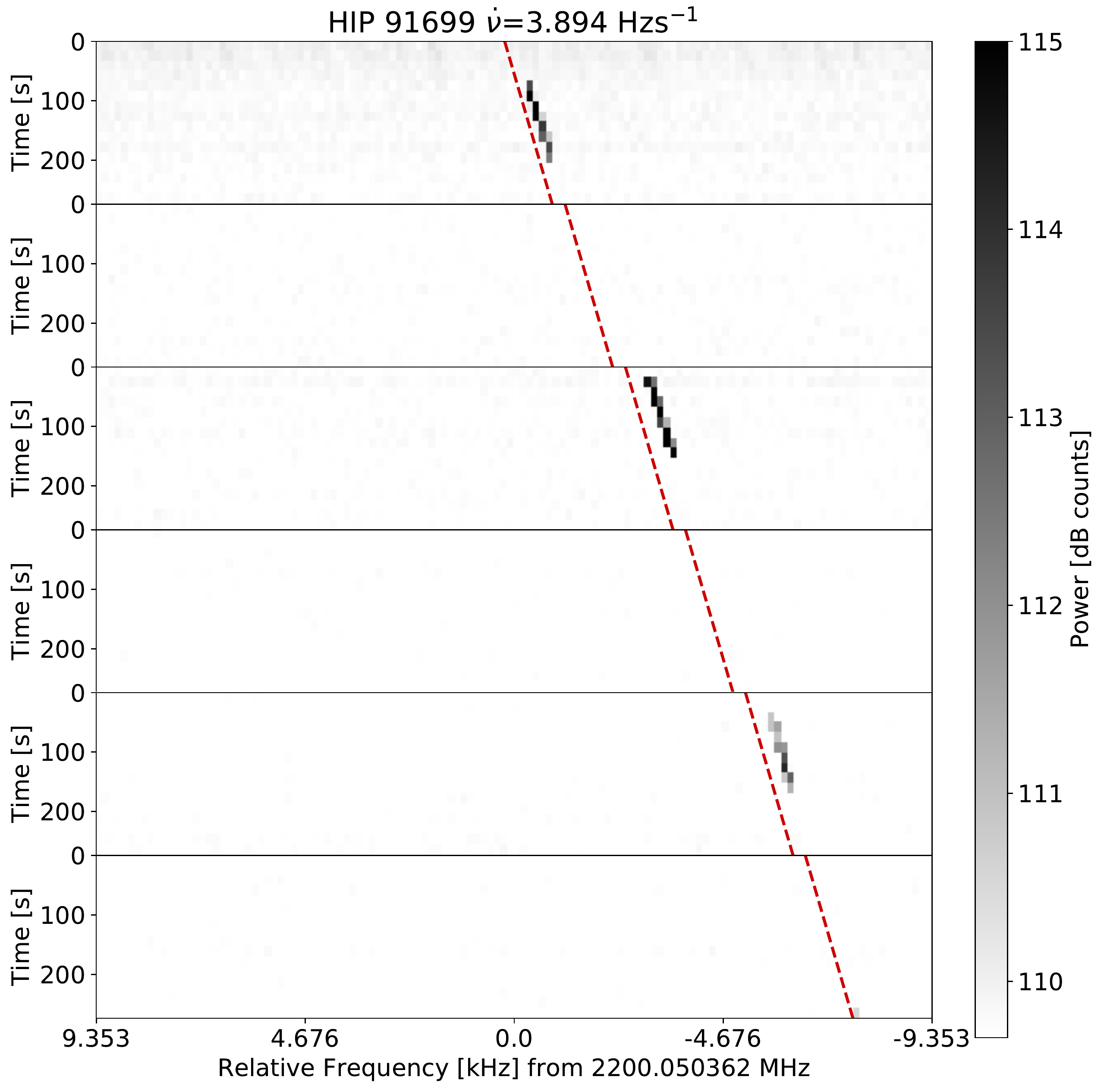}%
  \label{fig:S-event1}%
}\qquad
\subfloat[HIP22845]{%
  \includegraphics[width=0.45\columnwidth]{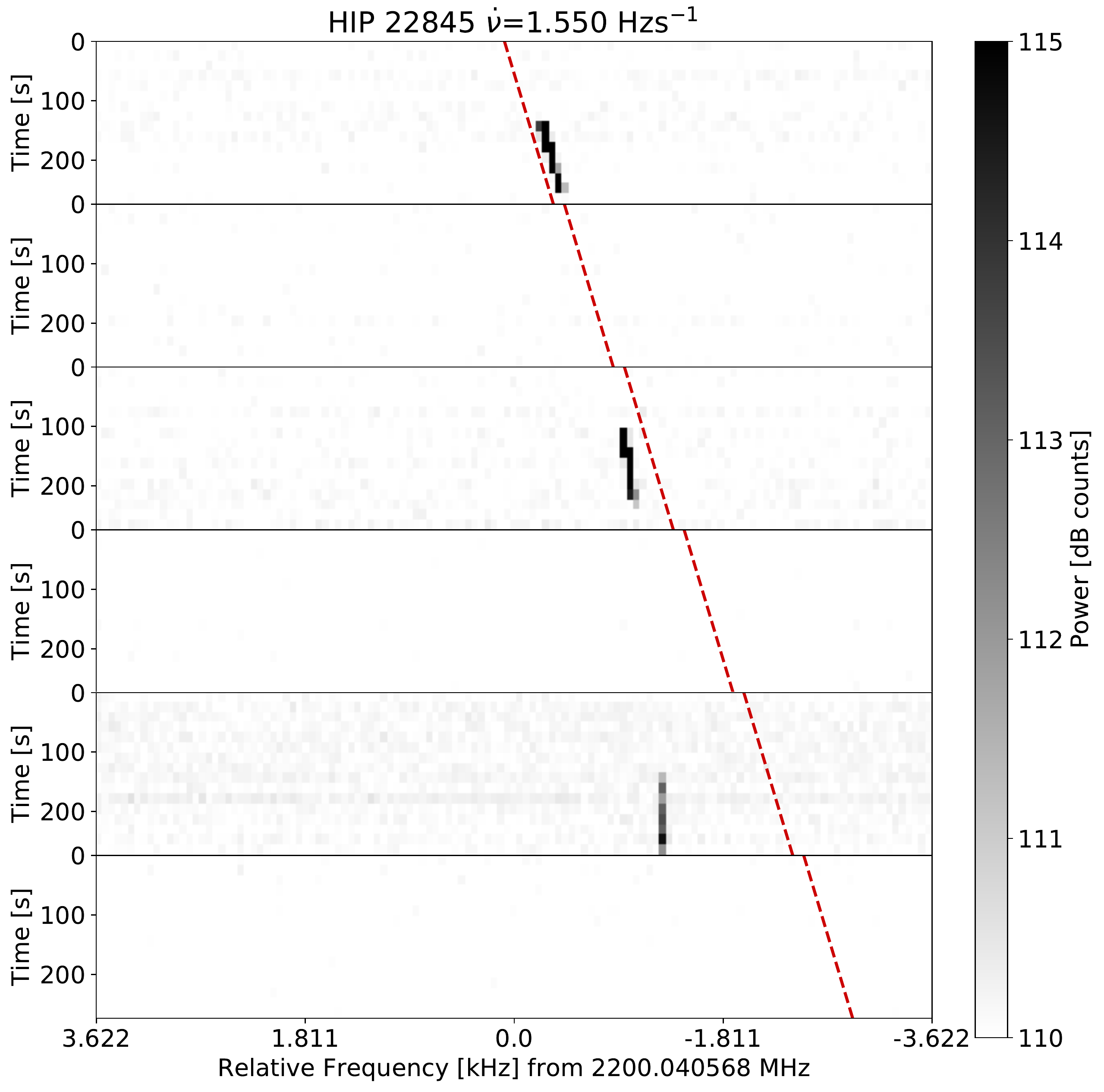}%
  \label{fig:S-event2}%
}

\subfloat[HIP44072]{%
  \includegraphics[width=0.45\columnwidth]{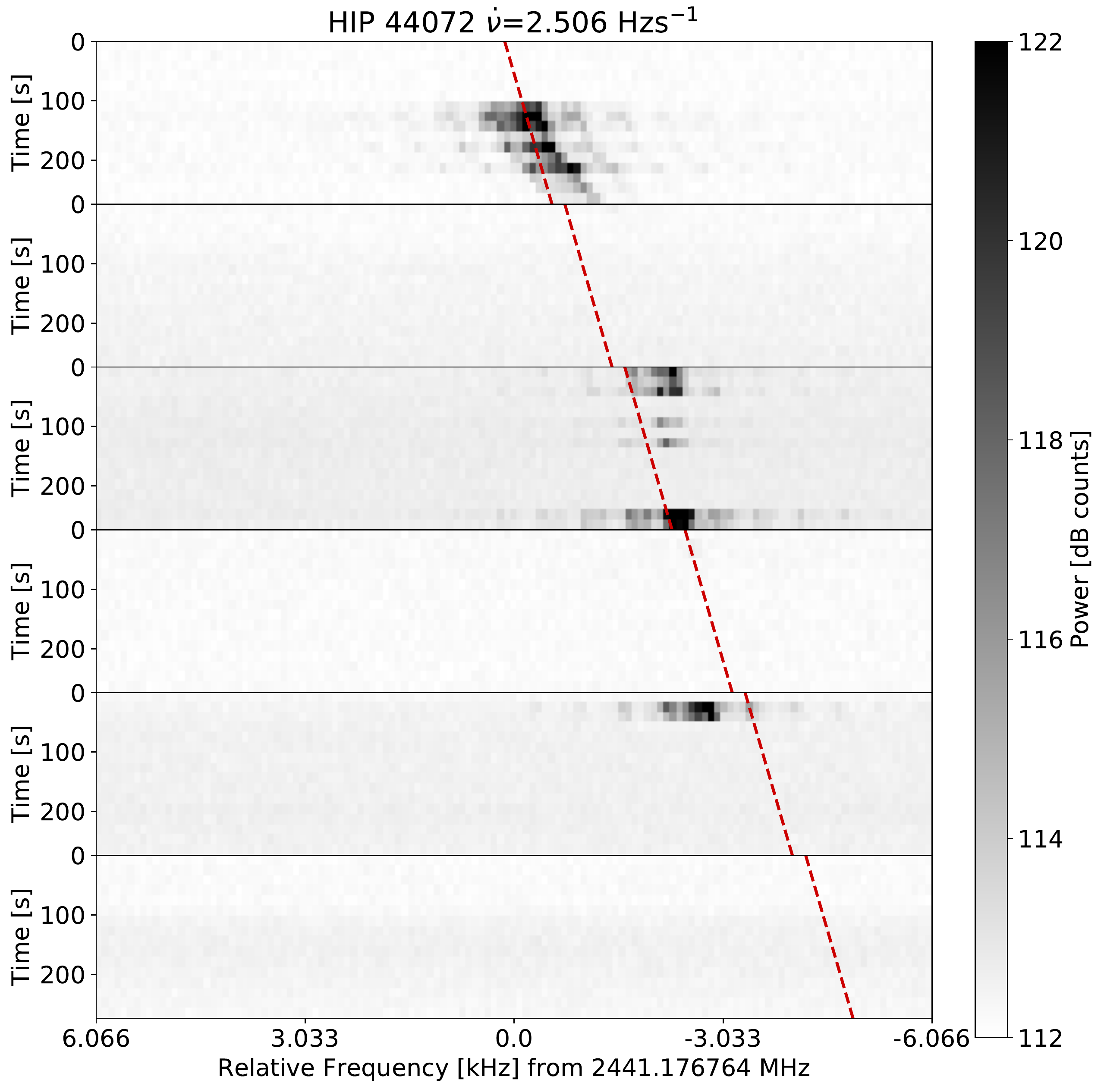}%
  \label{fig:S-event3}%
}\qquad
\subfloat[HIP109176]{%
  \includegraphics[width=0.45\columnwidth]{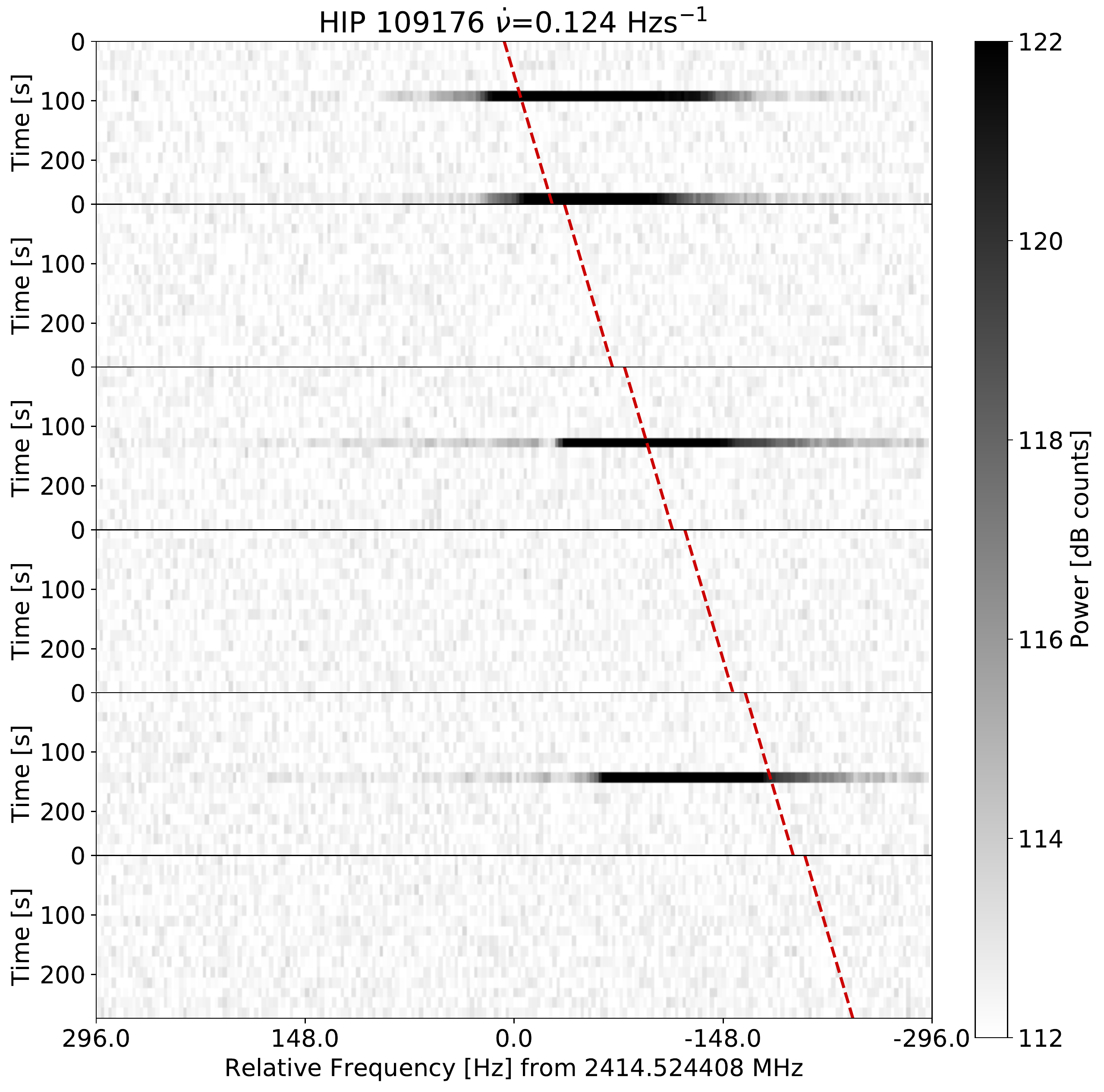}%
  \label{fig:S-event4}%
}
\caption{\label{fig:S-example-events}
Dynamic spectrum for selected \tseti events from GBT S-band observations. Each subfigure shows a full ABACAD cadence; each of the six panels represents ON and OFF source, consecutively.  The red dashed lines
show the drift rate as detected by the pipeline; a small frequency offset has been added for visualization.}
\end{figure*}

\begin{figure*}
\centering 
\subfloat[HIP113368]{%
  \includegraphics[width=0.45\columnwidth]{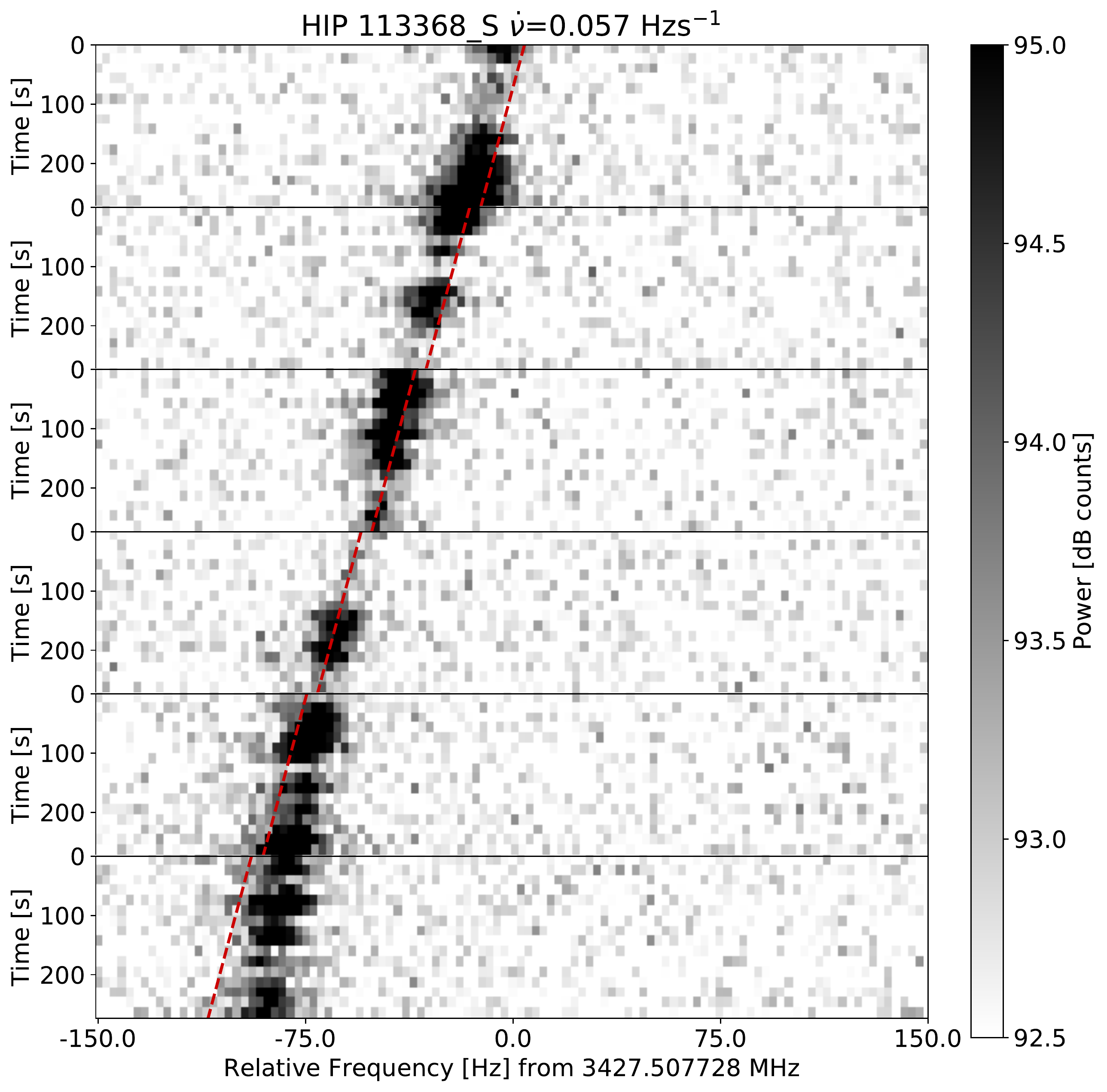}%
  \label{fig:pks-event1}%
}\qquad
\subfloat[HIP69454]{%
  \includegraphics[width=0.45\columnwidth]{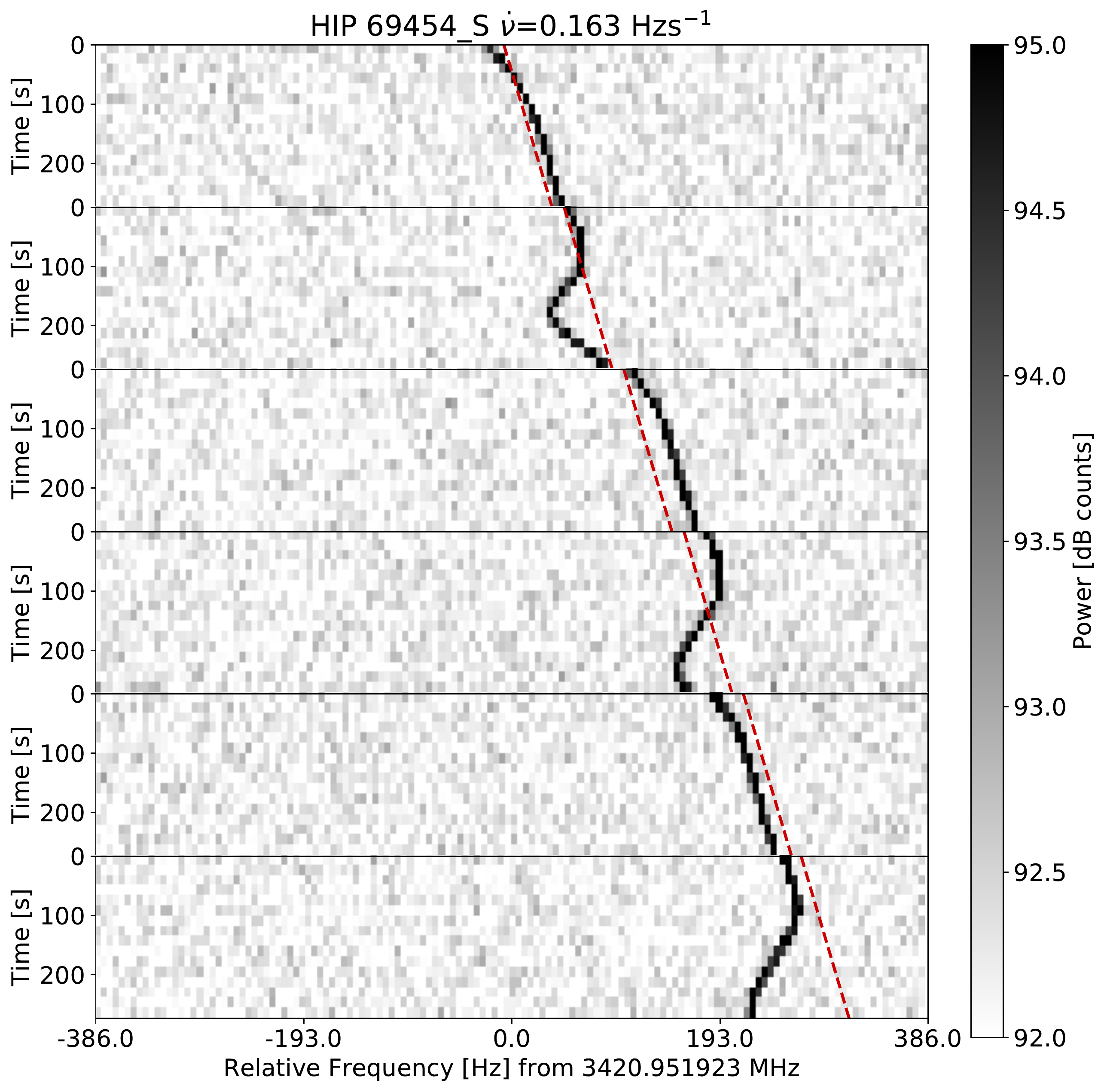}%
  \label{fig:pks-event2}%
}

\subfloat[HIP56452]{%
  \includegraphics[width=0.45\columnwidth]{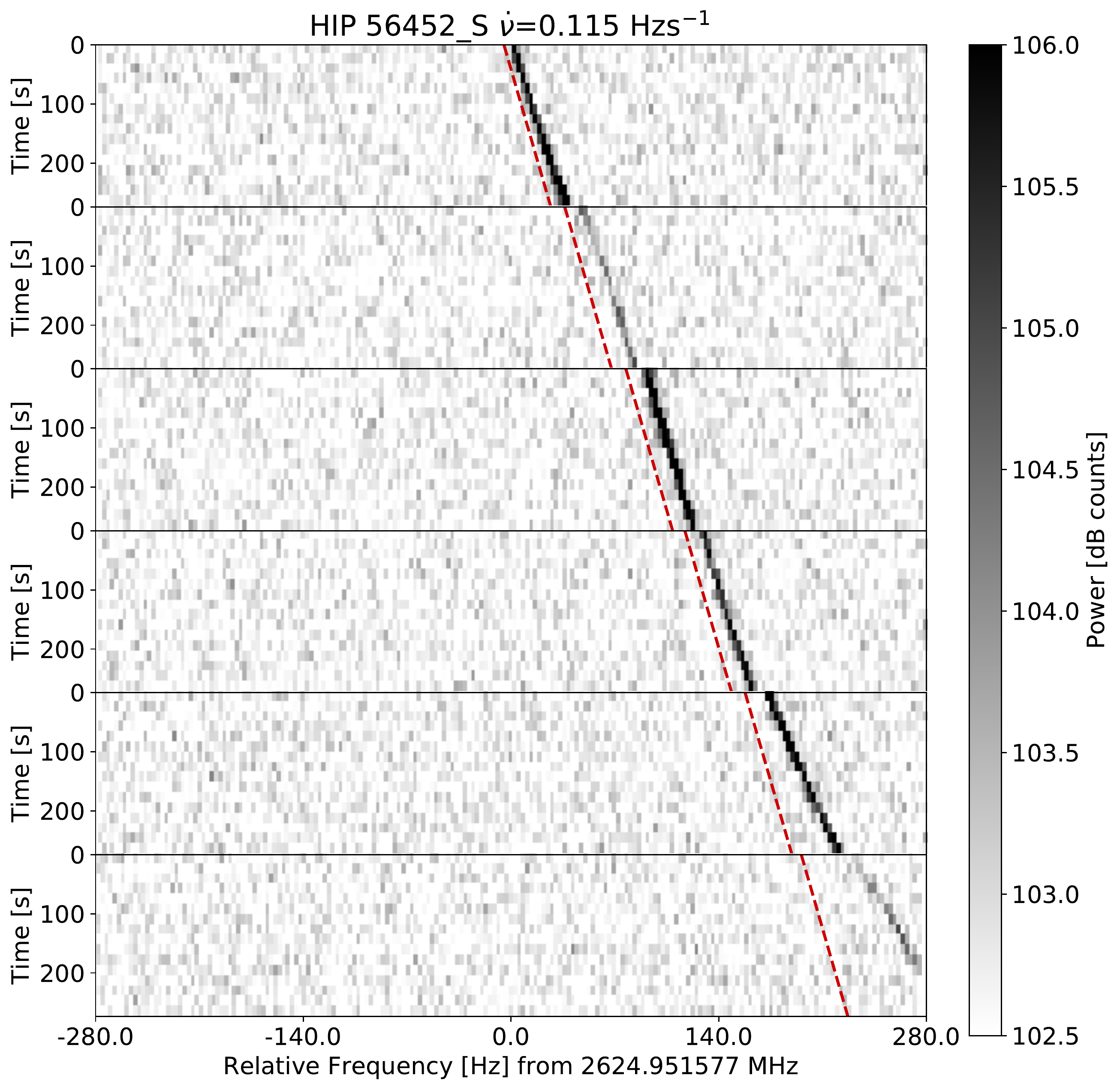}%
  \label{fig:pks-event3}%
}\qquad
\subfloat[HIP15330]{%
  \includegraphics[width=0.45\columnwidth]{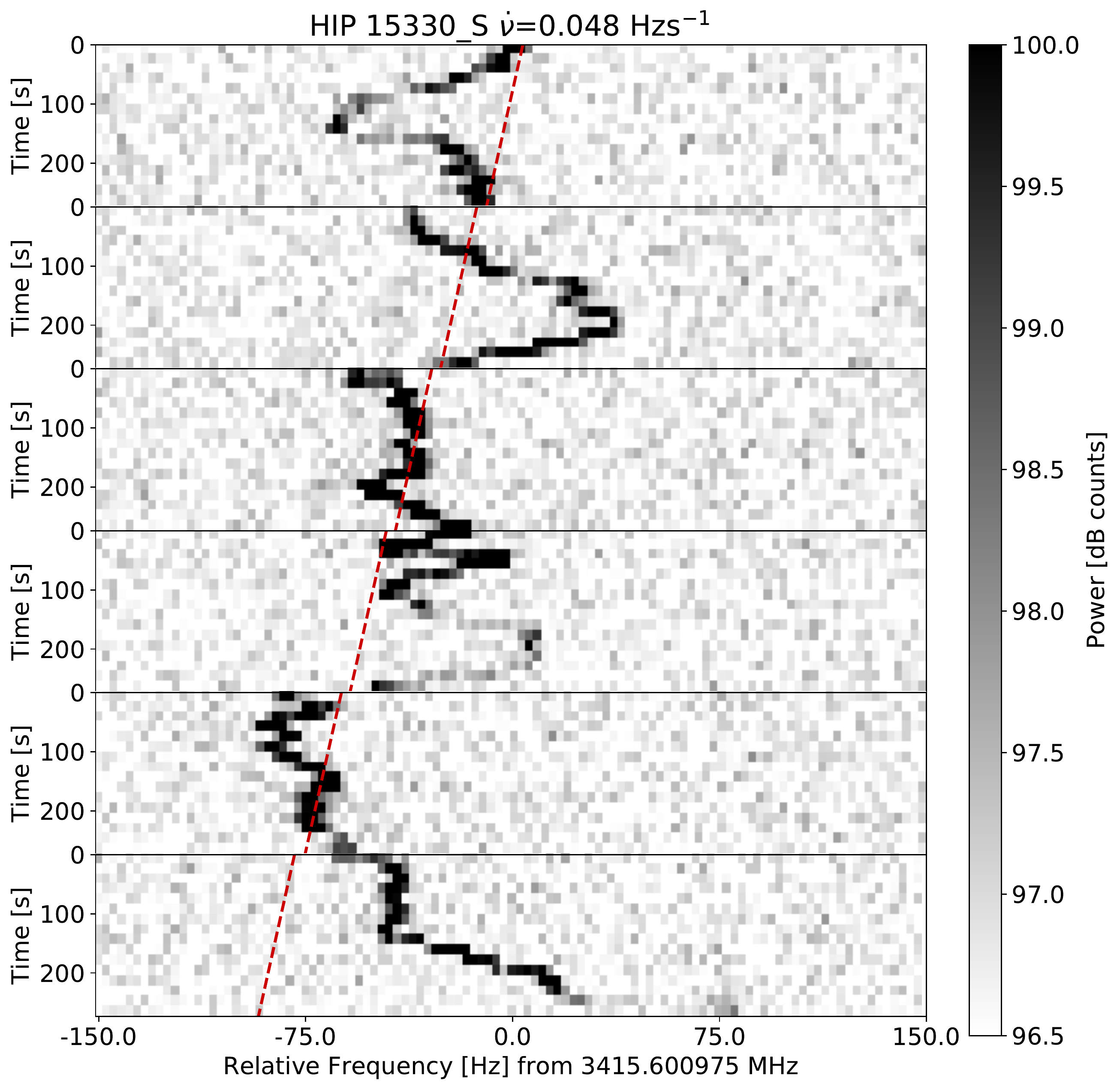}%
  \label{fig:pks-event4}%
}
\caption{\label{fig:pks-example-events}
Dynamic spectrum for selected \tseti events from Parkes 10-cm observations. Each subfigure shows a full ABABAB cadence; each of the six panels represents ON and OFF source, consecutively.  The red dashed lines
show the drift rate as detected by the pipeline; a small frequency offset has been added for visualization.}
\end{figure*}

A number of events passed our automated verification tests, but failed manual inspection. In this appendix, we discuss these events in further detail. Figures \ref{fig:L-example-events}, \ref{fig:S-example-events}, and \ref{fig:pks-example-events} show the four top-ranked events from each of the L-band, S-band and 10-cm datasets. \diff{These events are marked in \reffig{fig:freqs-all} with gold stars.}

\paragraph{At L-band (\reffig{fig:L-example-events})} Two of the most compelling events were from observations of HIP 54677 (\reffig{fig:L-event1} and \reffig{fig:L-event2}), appearing only in ON observations. These two signals are spaced $\sim$3\,MHz apart, have the same drift rate, similar bandwidth, and similar power levels. A similar event to \reffig{fig:L-event1} was detected $\sim$1\,kHz lower in an observation of HIP 103388; however this event was classified as RFI due to the presence of signal in OFF observations. In addition, RFI events were detected within 40\,Hz of the central frequency in \reffig{fig:L-event2}. Taken together, these RFI events indicate that the HIP 54677 events are also RFI. These events fall in the 1435--1525 MHz band used for aeronautical telemetry\footnote{\url{https://www.ntia.doc.gov/files/ntia/publications/.../1435.00-1525.00_01DEC15.pdf}}. 

As shown in \reffig{fig:L-event3}, \tseti detected an event only visible in ON pointings toward HIP 100064. A second drifting signal can be seen within the plotted band, however this is detected in both ON and OFF pointings. We reject this event as a similar pair of drifting signals are seen in observations of HIP 32423, within 30\,Hz of the central frequency of \reffig{fig:L-event3}; a similar pair is also seen in observations of HIP 21402, $\sim$1.5\,kHz above that of HIP 100064. 

\reffig{fig:L-event4} shows an event in the direction of HIP 1444. Events with similar signal bandwidths were detected $\sim$2.4\,kHz below (observations of HIP 56802), and $\sim$1.6\,kHz above (HIP 99572). However, these RFI events have differing drift rates, and do not exhibit the change of drift rate present in the HIP 1444 event. 

\paragraph{At S-band (\reffig{fig:S-example-events})} Narrowband signals with non-zero drift were detected above S/N threshold in observations of HIP\,91699  (\reffig{fig:S-event1}) and HIP\,22845 (\reffig{fig:S-event2}). These events have durations under 5 minutes, so appear to turn on and off during the observations. A total of 22 other events with similar drift rates and frequency-time structure were found within 2200.04--2200.5\,MHz, some of which appear in OFF pointings; as such, we identify the HIP 91699 and HIP 22845 events as RFI. The 2200--2290\,MHz band is used for spacecraft tracking and telemetry\footnote{\url{https://www.ntia.doc.gov/files/ntia/publications/.../2200.00-2290.00-01MAR14.pdf}}.

The HIP 44072 event (\reffig{fig:S-event3} displays complex structure in both time and frequency. Similar events were found in observations of HIP\,68030, HIP\,13402 and HIP\,113178; as such, we identify this as RFI. Similarly, HIP\,109716 is similar to events detected in HIP\,77655 pointings in which hits are visible in both ON and OFF pointings.

\paragraph{Parkes 10-cm receiver (\reffig{fig:pks-example-events})} Remarkably fewer events were detected at Parkes, with only 60 event groups passing automated verification. All of these signals are present in both ON and OFF observations, but were not detected above the required S/N of 10 in OFF observations. Given their frequency extent, most are likely associated with the National Broadband Network (NBN) that is known to operate at 3.4\,GHz.

\afterpage{\clearpage}

\afterpage{\clearpage}

\afterpage{\clearpage}

\newpage






%

\end{document}